\documentclass[twocolumn]{aastex631}

\usepackage{xcolor}
\usepackage{amsmath}
\usepackage{stix}
\usepackage{enumitem}
\usepackage{xspace}
\let\tablenum\relax
\usepackage[separate-uncertainty=true]{siunitx}

\newcommand{\teff}{\ensuremath{T_{\mathrm{eff}}}}

\graphicspath{{./}{figures/}}

\received{}
\revised{}
\accepted{}

\shorttitle{Fast-evolving Light Curves of VHS 1256 \MakeLowercase{b}}
\shortauthors{Zhou, Bowler et al.}

\begin{document}


\title{Roaring Storms in the Planetary-Mass Companion VHS 1256-1257 b:\\
Hubble Space Telescope Multi-epoch Monitoring Reveals Vigorous Evolution in an Ultra-cool Atmosphere}

\email{yifan.zhou@utexas.edu}
\author[0000-0003-2969-6040]{Yifan Zhou}
\altaffiliation{51 Pegasi b Fellow}
\affiliation{Department of Astronomy, The University of Texas at Austin, 2515 Speedway, Stop C1400, Austin, TX 78712, USA}

\author[0000-0003-2649-2288]{Brendan P. Bowler}
\affiliation{Department of Astronomy, The University of Texas at Austin, 2515 Speedway, Stop C1400, Austin, TX 78712, USA}

\author[0000-0003-3714-5855]{D\'aniel Apai}
\affiliation{Department of Astronomy/Steward Observatory, The University of Arizona, 933 N. Cherry Avenue, Tucson, AZ, 85721, USA}
\affiliation{Department of Planetary Science/Lunar and Planetary Laboratory, The University of Arizona, 1640 E. University Boulevard, Tucson, AZ, 85718, USA}

\author[0000-0003-3759-9080]{Tiffany Kataria}
\affiliation{Jet Propulsion Laboratory, California Institute of Technology, 4800 Oak Grove Drive, Pasadena, CA, USA}

\author[0000-0002-4404-0456]{Caroline V. Morley}
\affiliation{Department of Astronomy, The University of Texas at Austin, 2515 Speedway, Stop C1400, Austin, TX 78712, USA}

\author[0000-0002-6076-5967]{Marta L. Bryan}
\altaffiliation{NHFP Sagan Fellow}
\affiliation{Department of Astronomy,  University of California Berkeley, Berkeley, CA 94720-3411, USA}

\author[0000-0001-6098-3924]{Andrew J. Skemer}
\affiliation{Department of Astronomy, University of California Santa Cruz, 1156 High Street, Santa Cruz, CA USA}

\author[0000-0001-5578-1498]{Bj\"orn Benneke}
\affiliation{University of Montreal, Montreal, QC, H3T 1J4, Canada}

\newcommand{\vhs}{VHS~1256~b\xspace}

\begin{abstract}
  Photometric and spectral variability of brown dwarfs probes heterogeneous temperature and cloud distribution and traces the atmospheric circulation patterns. We present a new 42-hr Hubble Space Telescope (HST) Wide Field Camera 3 G141 spectral time series of VHS~1256-1257~b, a late L-type planetary-mass companion that has been shown to have one of the highest variability amplitudes among substellar objects. The light curve is rapidly evolving and best-fit by a combination of three sine waves with different periods and a linear trend. The amplitudes of the sine waves and the linear slope vary with wavelength, and the corresponding spectral variability patterns match the predictions by models invoking either heterogeneous clouds or thermal profile anomalies. Combining these observations with previous HST monitoring data, we find that the peak-to-valley flux difference is $33\pm2$\% with an even higher amplitude reaching 38\% in the $J$ band, the highest amplitude ever observed in a substellar object. The observed light curve can be explained by maps that are composed of zonal waves, spots, or a mixture of the two. Distinguishing the origin of rapid light curve evolution requires additional long-term  monitoring. Our findings underscore the essential role of atmospheric dynamics in shaping brown dwarf atmospheres and highlight VHS~1256-1257 b as one of the most favorable targets for studying atmospheres, clouds, and atmospheric circulation of planets and brown dwarfs.
\end{abstract}

\section{Introduction}

Brown dwarfs and wide-orbit (${>}100$~\si{au}) planetary mass companions (PMC) are often regarded as analogs to gas giant planets because they share similar effective temperatures, thermal profiles, and atmospheric compositions \citep[e.g.,][]{Beichman2014,Chabrier2014,Faherty2016,Bowler2017,Zhang2021}. Without bright host stars contaminating signals from the companions, observations of brown dwarfs and PMCs have delivered exquisite photometry, spectra, and time-series data that enable comprehensive atmospheric studies \citep[e.g.,][]{Vos2020,Miles2020,Best2021,Burningham2021,Apai2021,Vos2022}. These objects are valuable targets for understanding substellar atmospheric dynamics. Their high internal heat flux and fast rotation rates lead to intense thermal heating and strong Coriolis forces that define unique circulation regimes \citep{Zhang2014,Showman2019,Showman2020,Tan2020b,Tan2021}.

Driven by circulation, atmospheric structures form in substellar atmospheres and evolve over time. Heterogeneous distributions of condensate clouds and thermal profiles cause brightness and spectroscopic variability \citep[e.g.,][]{Morley2014,Robinson2014,Tremblin2020}, which has been found in a large number of brown dwarfs and a handful of PMCs \citep[e.g.,][]{Artigau2009,Radigan2012,Buenzli2014,Metchev2015,Zhou2016,Vos2017,Eriksson2019,Zhou2019,Bowler2020,Vos2020,Tannock2021,Vos2022}. Two types of variability patterns have emerged in observations, and they offer distinctive evidence for circulation shaping atmospheres. The first type is rotational modulations, manifesting as periodic and often (approximately) sinusoidal light curves \citep[e.g.,][]{Apai2013,Biller2017,Vos2017,Tannock2021}. This type of variability originates from a brown dwarf’s rotation transporting temperature and cloud anomalies in and out of the visible hemisphere, and hence the light curve constrains the object's rotation period despite the possible minor influence of differential rotation \citep[e.g.,][]{Metchev2015,Scholz2015,Zhou2016,Tannock2021}. The size and shape of the atmospheric structures determine the appearance of the light curve, and thus high-precision time series observations enable the reconstruction of top-of-atmospheric maps \citep{Apai2013,Karalidi2015,Karalidi2016}. The size of the retrieved spots constrains wind speeds in brown dwarf atmospheres based on the Rhines' length argument \citep{Rhines1975,Showman2002}. Combining periods measured at infrared and radio wavelengths, which trace the stratospheric and magnetospheric rotation rates, \citet{Allers2020} directly measured the wind speed of a brown dwarf. Amplitudes and phase offsets of rotational modulations often vary with wavelength \citep[e.g.,][]{Buenzli2012,Lew2016,Yang2016,Biller2017}, and the observed wavelength-dependence of the modulations supports a picture in which heterogeneous clouds are the primary source of variability in L/T transition dwarfs \citep[e.g.,][]{Apai2013,Buenzli2015,Lew2020b}.

The second type of variability is long-term and irregular light curve evolution. \citet{Artigau2009} and \citet{Radigan2012} discovered significant morphological differences between light curves separated by a few rotation periods. \citet{Metchev2015} found that the irregular changes were common among a large sample of brown dwarf light curves collected by the Spitzer Space Telescope. So far, the most comprehensive observational evidence of brown dwarf light curve evolution is from long time baseline brown dwarf monitoring campaigns conducted with Spitzer and TESS (by \citealt{Apai2017} and \citealt{Apai2021}, respectively). In these studies, light curves spanning over one hundred brown dwarf rotation periods exhibited a wealth of patterns, including sinusoids, beating of two similar frequencies, and irregular and non-periodic variations. The evolution of these patterns is hardly predictable: they may maintain a regular sine wave shape in a short time interval but then evolve dramatically over just a few rotations. \citet{Apai2017}  and \citet{Apai2021} found that the circular/elliptical spot models alone could not explain several evolving light curves. On the other hand, the planetary-scale wave model, which is supported by general circulation model (GCM) results \citep[e.g,][]{Showman2019,Tan2021}, fit observations much better.

Despite this progress, a fundamental question remains: what are the physical mechanisms that lead to brown dwarfs' heterogeneous atmospheres and long-term atmospheric evolution? This question has been explored in several studies modeling the circulation patterns in these atmospheres. A brown dwarf's vigorous convection can perturb its stratosphere and introduce inhomogeneity in temperature and thermal profile distributions \citep{Showman2012,Robinson2014}. Meanwhile, because the thermal profile of a brown dwarf intercepts the condensation curves of Fe and Si minerals, these species condense and coagulate into clouds \citep[e.g.,][]{Ackerman2001,Marley2002,Burrows2006a,Helling2008,Marley2013,Charnay2017,Gao2018}. Circulation regularizes cloud formation, carving clouds into patches \citep[e.g.,][]{Showman2012}. By tuning the GCM simulations to match the properties of typical brown dwarfs, \citet{Showman2019} found that zonal bands and jets are common outcomes. Furthermore, under conditions of short drag and radiative timescales,  planetary-scale waves induce long-term oscillations in a brown dwarf stratosphere. These oscillations are similar to those observed on Earth, Jupiter, and Saturn. Moreover, due to cloud radiative feedback, cloud thickness and the surrounding thermal profile can modulate spontaneously, introducing quasi-periodic variability that does not correlate with rotation \citep{Tan2018}. \citet{Tan2020b,Tan2021} further integrated cloud radiative feedback into GCMs and found that the vigorous atmospheric circulation triggered  and maintained by cloud formation can substantially impact the observable properties of brown dwarfs. The types of heterogeneous atmospheres can be identified by spectral variability \citep{Morley2014} and the circulation models can be probed through light curve morphology and variability timescales \citep[e.g.,][]{Zhang2014,Tan2021}. High-precision, multi-wavelength, and multi-epoch time-resolved observations provide the most thorough and direct way to test these models.

VHS J125060.192-125723.9 b (hereafter, \vhs), an L7 brown dwarf companion hosted by a late-M equal-mass binary \citep{Gauza2015,Stone2016}, is an excellent target for such observations. Immediately after its discovery, it became a target of interest for atmospheric studies because of its likely young age (${<}300$~Myr, \citealt{Gauza2015}), low surface gravity, red infrared colors, and spectral resemblance to directly imaged planets such as HR~8799~bcde. Follow-up observations have found that \vhs's atmosphere has thick condensate clouds \citep[e.g.,][]{Rich2016} and a low methane abundance that deviates from the expected value based on chemical equilibrium \citep{Miles2018}.
By comparing the luminosity of \vhs with evolutionary tracks, \citet{Dupuy2022} found the companion's mass to be $11.8\pm0.2 M_{\mathrm{Jup}}$  or $16\pm1M_{\mathrm{Jup}}$, depending on the model choices. It is also among the first targets to be observed by JWST as part of an Early Release Science program \citep{Hinkley2022}.
In time-resolved observations, \vhs has exhibited high-amplitude brightness and spectral variability \citep[e.g.,][]{Bowler2020,Zhou2020a}. The host binary stars show low-levels ($<0.3\%$) photometric modulations \citep{Miles-Paez2021}.

 The variability signals in \vhs revealed insightful information about its atmosphere. \citet{Bowler2020} conducted time-resolved observations of \vhs  with Hubble Space Telescope/Wide Field Camera 3 G141 grism in 2018. The six-orbit continuous monitoring (${\sim}9$~\si{\hour}) resulted in a light curve that spanned less than half of its rotation period. At this epoch, \vhs's \SIrange{1.1}{1.7}{\micro\meter} band-integrated light curve exhibited a $19.3\%$ peak-to-valley brightness change, the second highest ever found in a brown dwarf. Shortly after this discovery, a 38-hr Spitzer/IRAC 4.5~\micron{} campaign was conducted to recover the companion's full rotation period \citep{Zhou2020a}. This follow-up observation found a sinusoidal light curve that helped determine a precise period of $22.02\pm0.04$~\si{\hour}. \citet{Zhou2020a} interpreted this measurement as the rotation period of \vhs. The combined spectral variability in the \SIrange{1.1}{1.7}{\micro\meter} WFC3/G141 band and the \SI{4.5}{\micro\meter} Spitzer band agreed with predictions from partly cloudy models \citep{Morley2014}.

In this paper, we present a new set of HST/WFC3 spectroscopic light curves of \vhs collected in 2020. The new observations have a time baseline of \SI{42}{\hour}, covering approximately two rotation periods of \vhs. Combining this new data with the 2018 HST results probes the companion's long-term changes on a timescale of nearly ${\sim}900$ rotation periods, offering a detailed view of \vhs's stormy atmosphere. This paper is organized as follows: we describe the observations and the data reduction method in \S\ref{sec:obs}, present the immediate results from the 2020 campaign in \S\ref{sec:results}, and analyze the long-term changes between the 2018 and 2020 epochs in \S\ref{sec:long-term}. Then in \S\ref{sec:discussion}, we discuss the rotation period measurement, the atmospheric circulation patterns, and the unusually high variability amplitude of \vhs. We summarize our results and conclude in \S\ref{sec:conclusions}.

\section{Observations and Data Reduction}
\label{sec:obs}
\newcommand{\rotationperiod}{\ensuremath{P_{\mathrm{rot}}}\xspace}

\begin{figure*}[ht]
  \centering
  \includegraphics[height=0.3\textwidth]{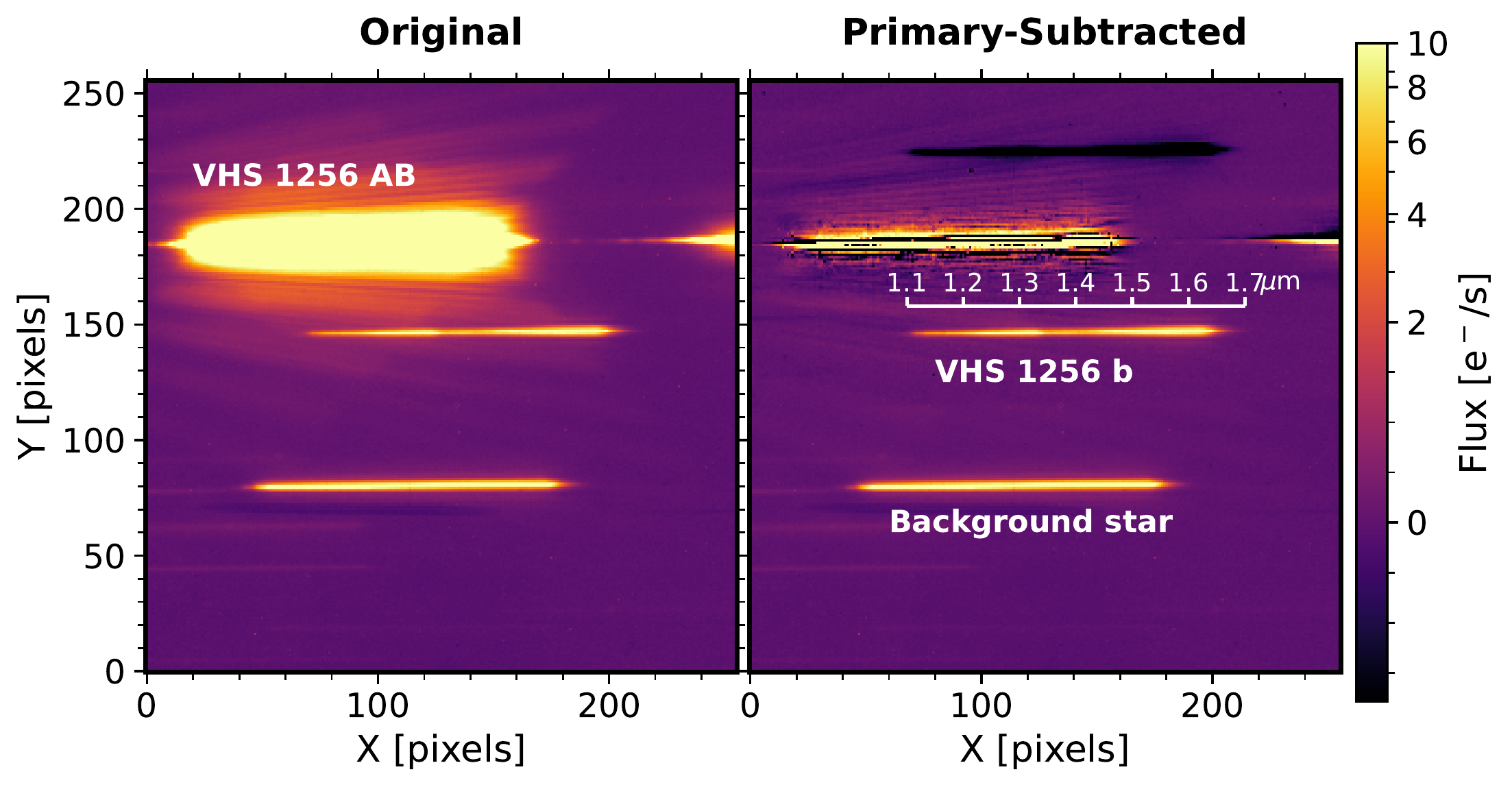}
  \includegraphics[height=0.3\textwidth]{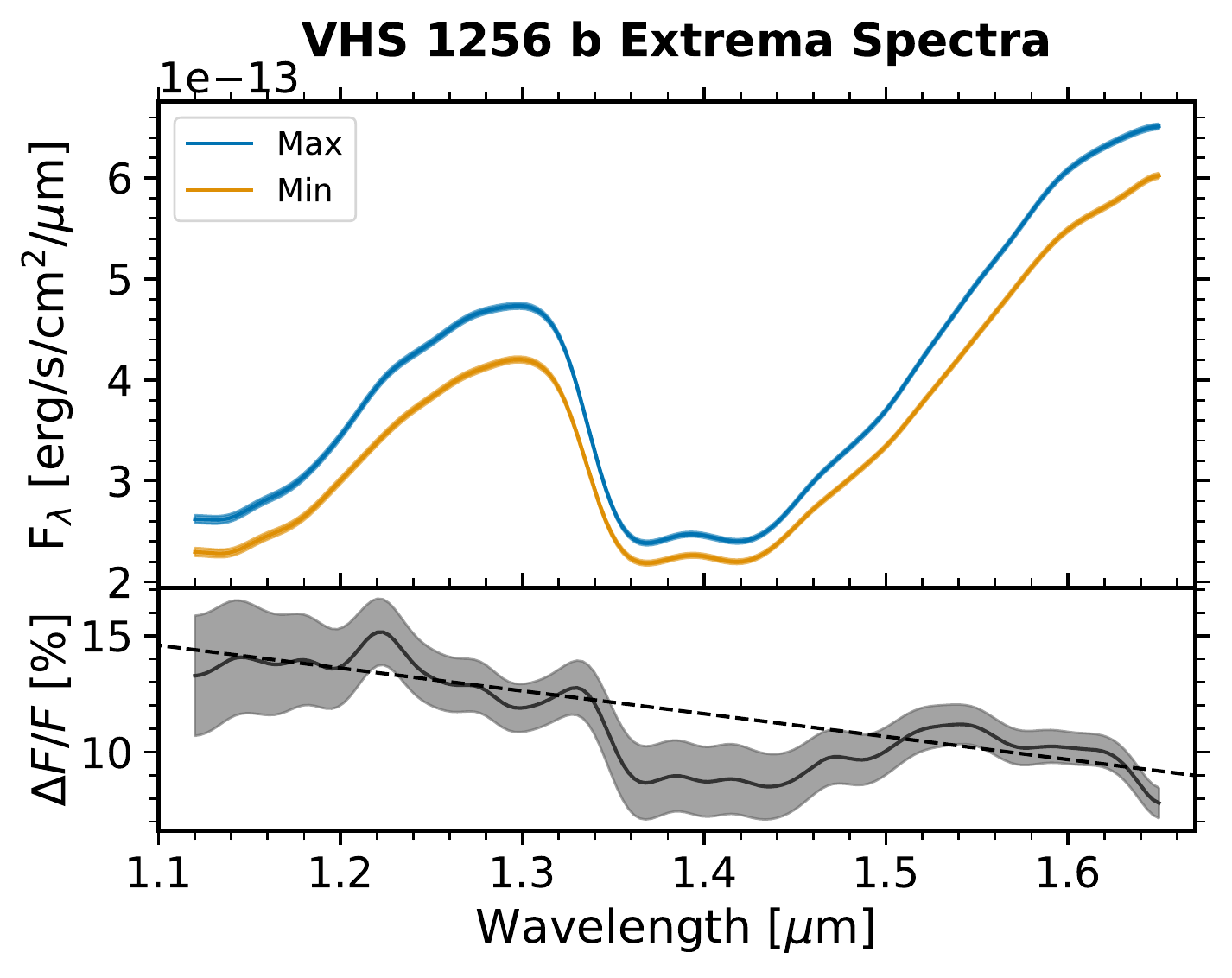}
    \caption{The median-combined original (left) and primary-subtracted (middle) images of the Epoch 2020 observations, and the extracted spectra of \vhs (right) at maximum and minimum brightness. In the left panel, from top to bottom, the three brightest spectral traces are VHS~1256~AB (the host binary), \vhs (the substellar companion), and a background star (2MASS J12560179${-}$1257390). The \vhs trace is moderately contaminated by the primary PSF. We mitigate the contamination by subtracting an empirically derived PSF model, and the primary-subtraction result is shown in the middle panel.  The right panel shows the extracted  spectra of \vhs at its brightest (blue) and faintest (yellow) phases. Their relative difference and the uncertainty are shown as a black line and a gray shaded region in the bottom subplot of the right panel.}
  \label{fig:specImage}
\end{figure*}

We conducted spectroscopic monitoring of \vhs using the Hubble Space Telescope Wide Field Camera 3 (HST/WFC3) IR Channel for fifteen orbits from UT 2020-05-26 09:27:17 to 2020-05-28 03:29:24  (Program ID: GO-16036, PI: Zhou; hereafter, we refer to this program as ``Epoch 2020''). Prior to this program, our target was observed with the same instrument for six continuous orbits from UT 2018-03-05 16:02:30 to 2018-03-06 00:42:47 (Program ID: GO-15197, PI: Bowler, results are published in \citealt{Bowler2020}; hereafter, we refer to this program as ``Epoch 2018'').

The overall design of the two observing campaigns are identical. Each orbit started with two to four direct-imaging exposures in the F132N filter for wavelength calibration. Eleven \SI{223}{s} spectroscopic frames in the G141 grism then followed. The spectrograph has a spectral resolution of $R\sim130$ at \SI{1.4}{\micro\meter} and a wavelength range spanning from \SIrange{1.12}{1.65}{\micro\meter}. The median combined spectral image obtained in the Epoch 2020 observations is shown in the left panel of Figure~\ref{fig:specImage} with the spectral trace of \vhs in the middle and two nearby sources above (VHS~1256${-}$1257~AB) and below (background star 2MASS J12560179${-}$1257390).

In two respects, the Epoch 2020 campaign differs from the Epoch 2018 one. First, the 2020 campaign has a longer time baseline. The purpose was primarily to monitor the brown dwarf companion for multiple rotations. For optimal time coverage and scheduling flexibility, we divided the observations into four segments. The first segment contained nine consecutive orbits tracking high-cadence (with a rate of \SI{242}{\second} per frame) variability for \SI{13.4}{\hour}, approximately 60\% of \vhs's \SI{22.0}{\hour} rotation period (\rotationperiod). After a gap of five orbits, three two-orbit segments followed with gaps of two and four orbits separating them. These three short segments significantly extended the overall time baseline. As a result, the entire campaign spanned \SI{42.0}{\hour} or 1.91 \rotationperiod of \vhs.

Second, the position angles (PAs) of HST are different between the two campaigns: $\mathrm{PA_{2020}}=299.0^{\circ}$ and $\mathrm{PA_{2018}}=143.1^{\circ}$\footnote{Following the definition used in the \href{https://hst-docs.stsci.edu/wfc3ihb}{WFC3 Instrument Handbook} and FITS file headers, these angles refer to the PA of HST's V3 axis.}. This difference is due to the fact that the allowed telescope PAs were set as ranges instead of fixed values. 

The PA orientation of the Epoch 2020 observations was less optimal than that in the Epoch 2018 observations, resulting in a smaller projected separation between the spectral traces of \vhs and VHS~1256~AB on the detector (38 pixels in Epoch 2020 vs. 57 pixels in Epoch 2018). The proximity between the two traces caused modest contamination from the host binary's point spread function at the position of the companion (Figure~\ref{fig:specImage}). Therefore, primary subtraction is necessary for accurately extracting spectra from the Epoch 2020 data.

We adopt a ``flip-and-subtract'' approach to remove contaminating flux. It includes four steps:
\begin{enumerate}
\item Construct an empirical point spread function (PSF)  template by median-combining all sky-subtracted spectroscopic images.
\item Flip the template upside down (i.e., mirror the image with respect to the $x$-axis); shift and tilt the flipped template to align the template spectral trace with the one in observed images; and linearly scale the template to match the flux with observed spectral trace.
\item Subtract the flipped template from the target image and calculate the sum of squared residuals in an optimization region. This is chosen to be a $15\times140$ rectangle 20 pixels above \vhs's trace.
\item Optimize the shift distance, tilt angle, and scaling factor by minimizing the sum of squared residuals. After the optimal values were found, use them to derive the final primary-subtracted images.
\end{enumerate}
The middle panel in Figure~\ref{fig:specImage} shows an example of primary-subtracted images. After subtraction, the contamination, measured as the average flux in the optimization region, is below the average sky background and hence does not cause significant uncertainties in photometry. We proceeded to use the primary-subtracted images to extract the spectra of \vhs.

The remaining data reduction procedures are identical to those detailed in \citet{Bowler2020}.
In the following Section, we discuss \vhs's atmospheric properties based on its spectra and light curves presented in Figures \ref{fig:specImage} and \ref{fig:lightcurves}, respectively.

\section{Results}
\label{sec:results}
\subsection{Light Curve Analysis}
\label{sec:lc-fit}

\begin{figure}[t]
  \centering
  \includegraphics[width=\columnwidth]{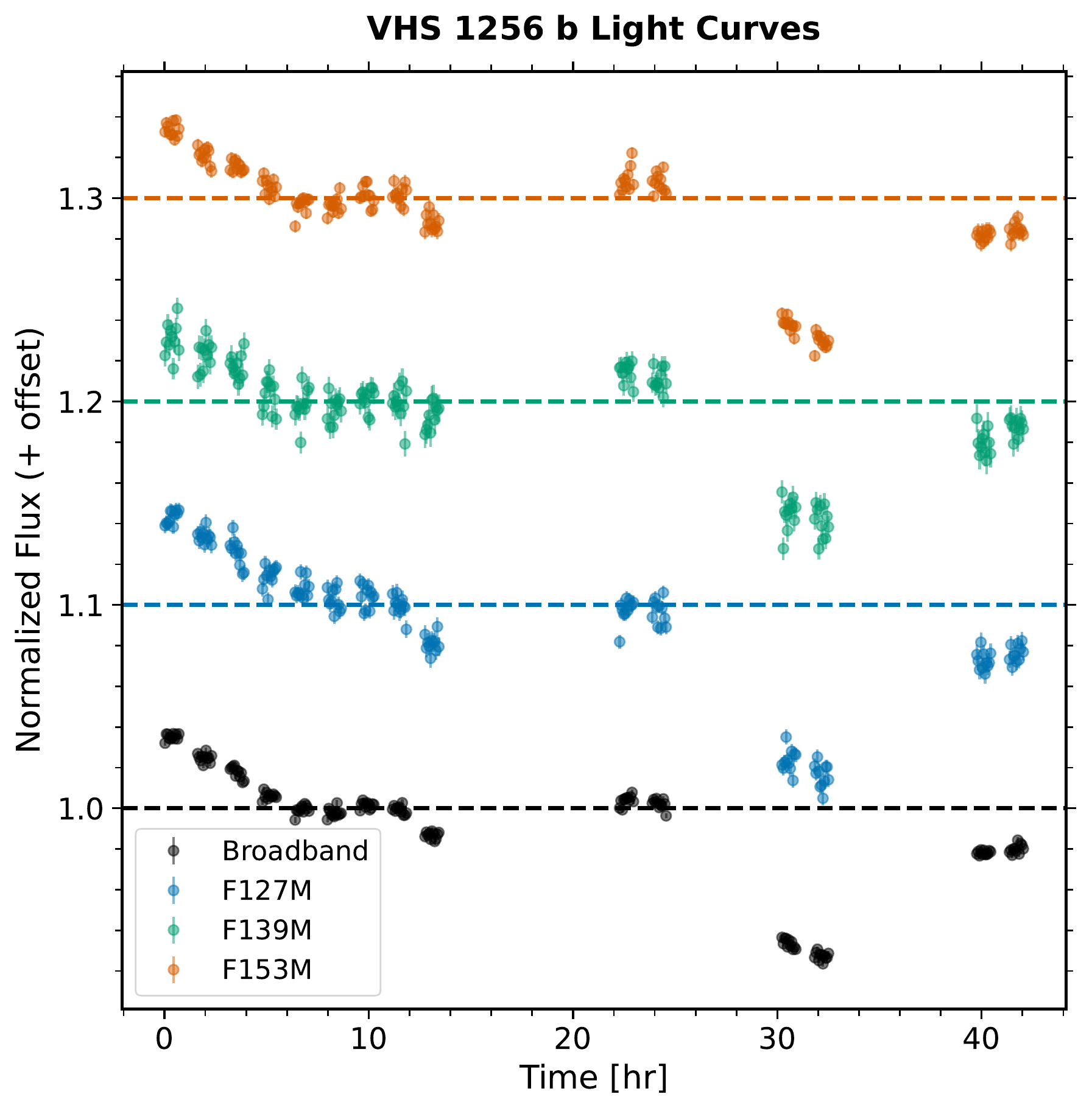}
  \caption{Light curves of \vhs integrated in four representative bandpasses obtained in the 2020 observations: the G141 broadband (black), the F127M filter (blue), the F139M filter (green), and the F153M filter (red). High-amplitude brightness modulations are detected in every light curve.}
  \label{fig:lightcurves}
\end{figure}

\begin{figure*}[!t]
  \centering
  \includegraphics[width=0.49\textwidth]{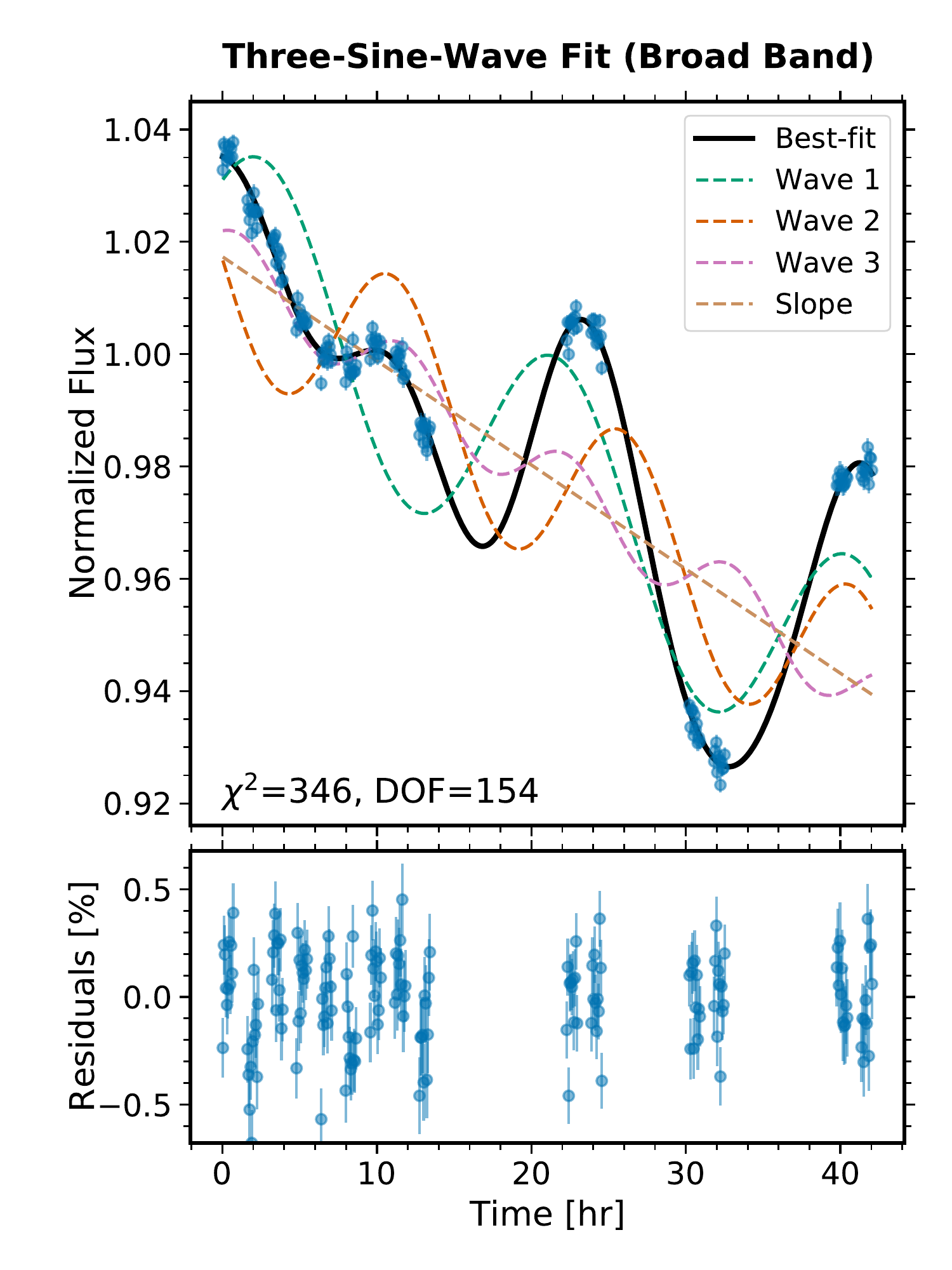}
  \includegraphics[width=0.49\textwidth]{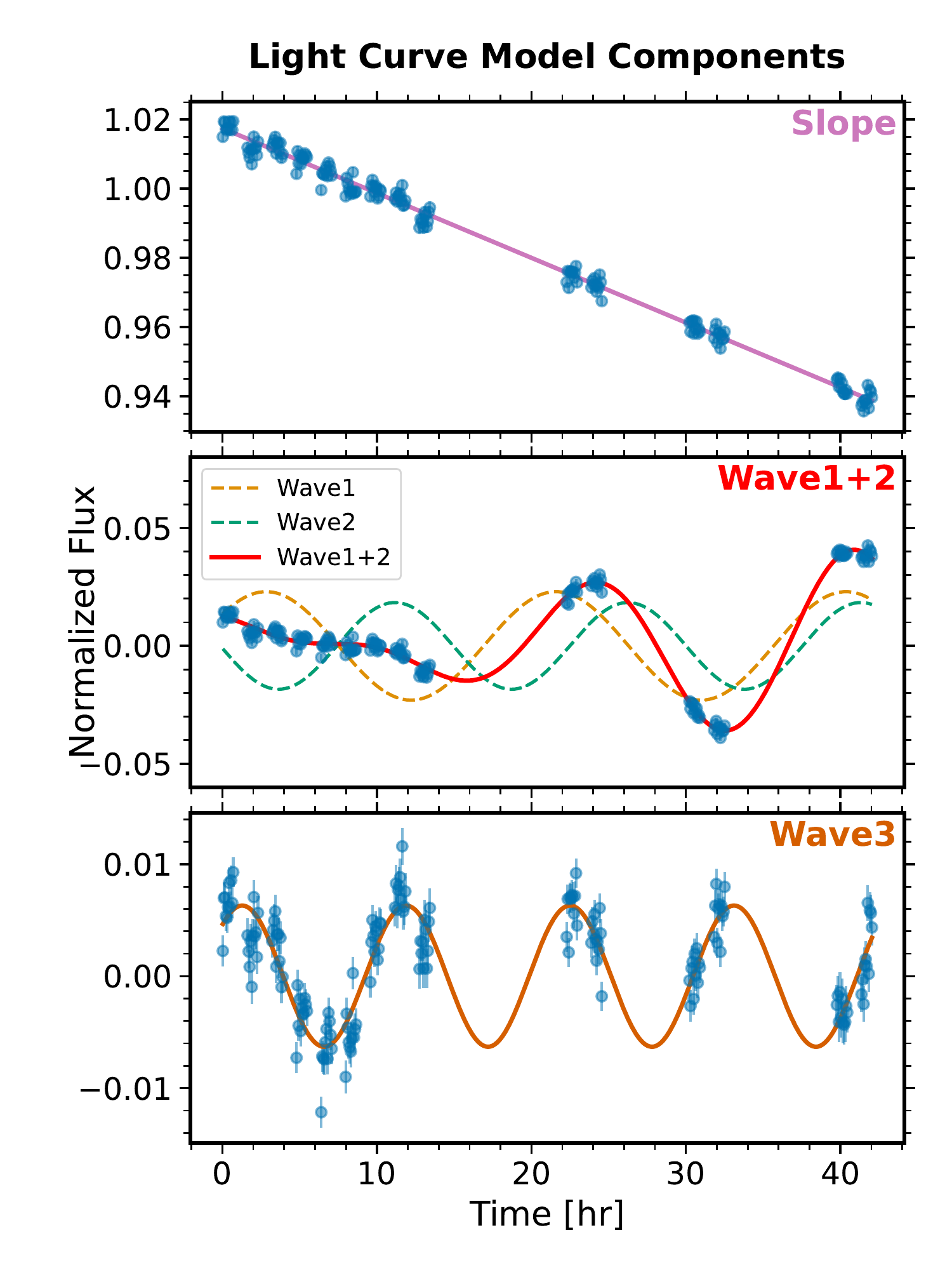}
  \caption{Fitting the multiple-sinusoidal model to the broadband light curve. The left panel highlights the excellent agreement between the observed broadband light curve (blue dots) and the best-fitting  model (black solid line). Individual model components (the three sine waves and the linear trend) are shown as color-dashed lines. In the right panel, we isolate these components to illustrate their importance to the good fit. The upper, middle, and lower panels feature the linear slope, Waves 1+2, and Wave 3, respectively. In each panel, model components other than the one being highlighted are subtracted from both the observed and model light curves. The middle-right panel demonstrates the combined effect of two sine waves that have best-fitting periods of 18.8~hr and 15.1~hr. In this panel, individual waves are plotted in the dashed lines and their sum is in the solid red line.}
  \label{fig:lcfit}
\end{figure*}

\begin{figure*}[!t]
  \centering
  \includegraphics[width=1.0\textwidth]{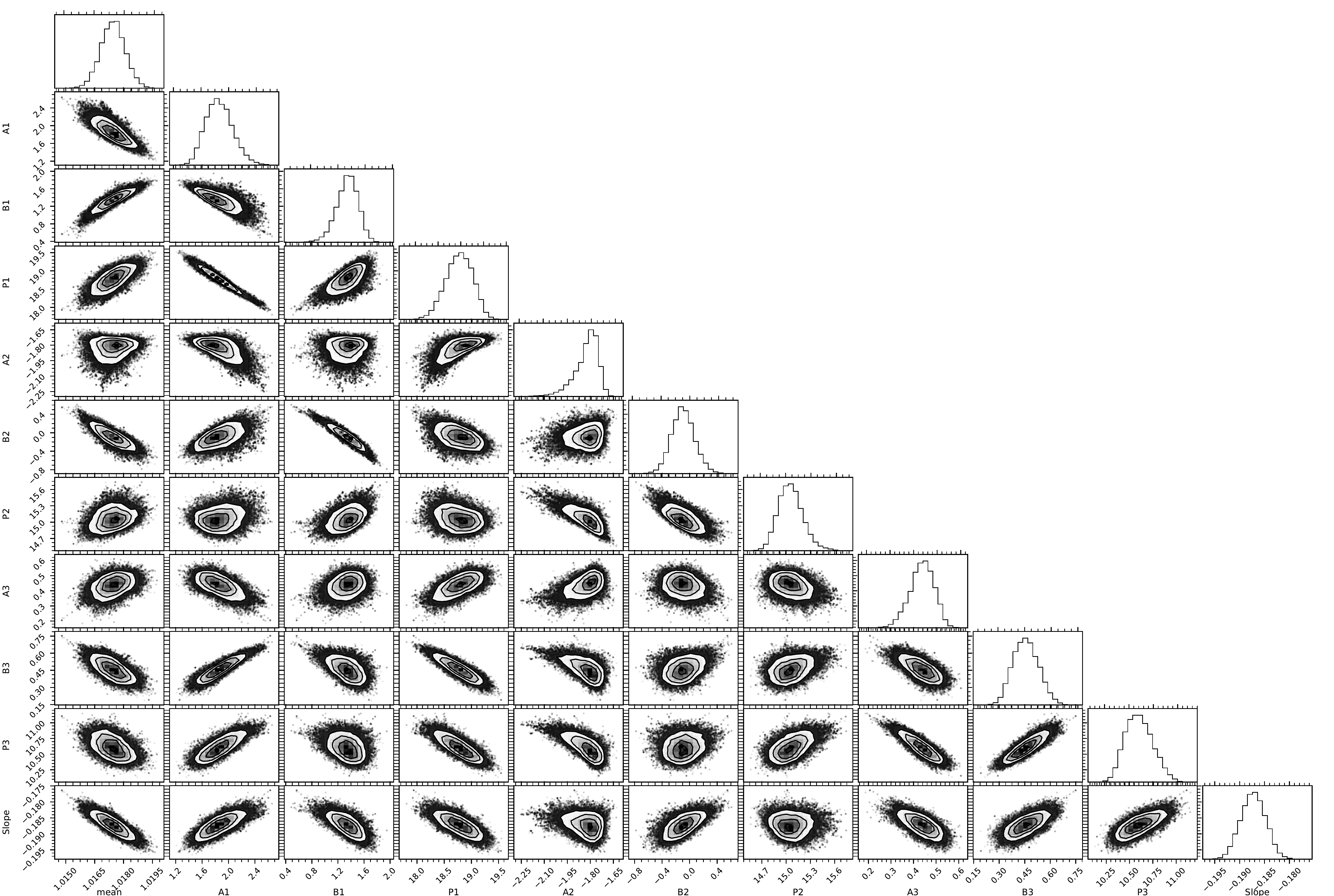}
  \caption{The posterior distributions of the three-sine-wave parameters in a corner plot.}
  \label{fig:lcfit_corner}
\end{figure*}

\begin{deluxetable*}{ccllllllllllll}
  \tablecolumns{14}
  \tablewidth{0pt}
  \tablecaption{$\chi^{2}$ and BIC values in Light Curve Fittings} \label{tab:lcfit}
  \tablehead{
    \colhead{$N$}
    & \colhead{DOF} 
    & \multicolumn{3}{c}{Broadband}
    & \multicolumn{3}{c}{F127M}
    & \multicolumn{3}{c}{F139M}
    & \multicolumn{3}{c}{F153M}\\
    && $\chi^{2}$ & BIC & $\Delta$BIC& $\chi^{2}$ & BIC & $\Delta$BIC& $\chi^{2}$ & BIC & $\Delta$BIC&
    $\chi^{2}$ & BIC & $\Delta$BIC
  }
  \startdata
  1 & 160 & 5580 & -26602 &      & 1562 & -4103 &     & 442.8 & -1644 &      & 1444 & -6364 & \\
  2 & 157 & 686.7 & -31480 & 4878 & 335.9  & -5314 & 1211 & 260.9 & -1811 & 167 & 381  & -7411 & 1047\\
  3 & 154 & 345.9 & -31806 & 326  & 280.0  & -5354 & 40   & 229.9 & -1826 & 15  & 291  & -7487 & 76\\
  4 & 151 & 291.2 & -31845 & 41   & 272.3  & -5346 & -8   & 229.4 & -1812 & -14 & 262  & -7500 & 13 \\
  \hline
   \multicolumn{2}{c}{Favored $N$} & \multicolumn{3}{c}{4} & \multicolumn{3}{c}{3} & \multicolumn{3}{c}{3} & \multicolumn{3}{c}{4}
  \enddata
  \tablenotetext{a}{BIC values are relative to those for the best-fitting linear trend.}
  \tablenotetext{b}{$\Delta$BIC is the BIC decrease due to adding one sinusoid. When $\Delta\mbox{BIC} > 10$, the more complex model is favored.}
\end{deluxetable*}

Figure~\ref{fig:lightcurves} shows the normalized light curves of \vhs during the Epoch 2020 observations in four representative bands: the G141 bandpass (\SIrange{1.12}{1.65}{\micro\meter}), the F127M filter (centered at \vhs's $J$-band peak, $\lambda=\SI{1.27}{\micro\meter}$, $\mathrm{FWHM}=\SI{0.069}{\micro\meter}$), the F139M filter (sensitive to water absorption, $\lambda=\SI{1.38}{\micro\meter}$, $\mathrm{FWHM}=\SI{0.065}{\micro\meter}$), and the F153M filter (continuum emission on the red side of the water absorption band $\lambda=\SI{1.53}{\micro\meter}$, $\mathrm{FWHM}=\SI{0.069}{\micro\meter}$). High-amplitude modulations are present in all light curves. Unlike the Epoch 2018 light curves that were well-fit by single sinusoids \citep{Bowler2020,Zhou2020a}, the Epoch 2020 light curves exhibit irregular structures.

We seek the simplest model that recovers the four band-integrated light curves. Following Apai et al. (\citeyear{Apai2017} and \citeyear{Apai2021}), we construct a model by summing multiple sinusoids and set the periods of all sinusoids to be free parameters. When the periods are harmonics, this model becomes a truncated Fourier series. In addition to sinusoids, we include a linear term to encapsulate the long-term atmospheric changes of which the timescales significantly exceed the observing window \citep[e.g.,][]{Biller2017}.

An $N$-component multi-sinusoidal model is expressed as
\begin{equation}
  \label{eq:3-2}
  F(t) = C_{0} + C_{1}\,t + \sum_{i}^{N}\Bigl( A_{i}\sin(2\pi t/P_{i}) + B_{i}\cos(2\pi t/P_{i}) \Bigr),
\end{equation}
where $A_{i}$ and $B_{i}$ are the amplitudes of the $i$th order sine and cosine components, respectively. $C_{0}$ is a normalization factor, and $C_{1}t$ is the linear trend. There are $3N+2$ free parameters. 

We incrementally increase $N$, and use the maximum likelihood method to find the best-fitting $A_{i}$, $B_{i}$, and $P_{i}$. The likelihood function is defined as:
\begin{equation}
  \label{eq:4}
  \mathcal{L} = \prod_{j}^{n}\frac{1}{\sqrt{2 \pi \sigma_{j}^{2}}} \exp\Bigl(-\frac{\bigl(f_{j}-F(t_{j})\bigr)^{2}}{2\sigma_{j}^{2}}\Bigr),
\end{equation}
where $f_{j}$, $\sigma_{j}$, and $t_{j}$ are the normalized flux density, the uncertainty, and the timestamp of the $j$th data point. We assume uninformative (uniform) priors for the free parameters and fit the model by sampling the posterior probability function using  Markov Chain Monte Carlo (implemented by \texttt{emcee}, \citealt{Foreman-Mackey2012}). For model selection, the Bayesian Information Criterion (BIC) is used. It is derived as
\begin{equation}
  \label{eq:6}
  \mathrm{BIC} = \chi^{2} + k\,\ln(n),
\end{equation}
in which $\chi^{2}$, $k$, and $n$ are the nominal chi-square values between the model and the data, the number of free parameters, and the number of data points, respectively. The fitting statistics are listed in Table~\ref{tab:lcfit}.

We determine the truncation order $N$ based on the BIC values with $\Delta\mbox{BIC} > 10$ as the threshold for favoring a more complex model \citep[e.g.][]{Kass1995}.  For all four light curves, the $N=3$ models are favored over any of the simpler models. For the broadband and F153M light curves, the even more complex $N=4$ model is preferred. However, because the $N=4$ model is not supported in all cases, we select the simpler $N=3$ model for the subsequent analysis.

Our best-fitting model is presented in Figure~\ref{fig:lcfit} and the posterior distributions of the model parameters are shown in Figure~\ref{fig:lcfit_corner}.
Based on the broadband light curve fit, the three sine waves have periods of: $18.8\pm0.2$~\si{\hour} (Wave 1), $15.1\pm0.2$~\si{\hour} (Wave 2), and $10.6\pm0.1$~\si{\hour} (Wave 3). The peak-to-peak amplitudes of the three waves are $5.8\pm0.8$\si{\percent}, $4.6\pm0.7$\si{\percent}, and $1.4\pm0.1$\si{\percent}. Waves 1 and 2 form a beating pattern and Wave 3 has a period close to one half of the 22~hr period best-fit to the Spitzer 4.5~\si{\micro\meter} light curve. In Figure~\ref{fig:lcfit}, we decompose the model and visualize how each component contributes to the observed light curve evolution. In the first half, Waves 1 and 2 are in nearly opposite phases and they cancel each other out. In the same segment, the modulations are mostly from Wave 3 and the linear trend. In the second half, Waves 1 and 2 start to align in phase and jointly increase the total modulation amplitude.
Notably, the periods of the first two waves are significantly shorter than the one best-fit to \vhs's Spitzer light curve \citep[\SI{22.04 }{hr},][]{Zhou2020a}. We discuss the discrepancy of the period measurements in \S\ref{sec:period-discussion}.

Our light curves can be equally well fit by a Fourier series truncated at the fourth order with a base-order period of 41.4~hr combined with a linear trend. This truncated Fourier series has the same number of free parameters as our best-fitting model. For the broadband light curve, the four Fourier components have peak-to-peak amplitudes of 1.74\%, 4.14\%, 2.92\%, and 0.90\% for orders 1 to 4, respectively. Components 2 to 4 ($P_{2}{=}20.7$~hr, $P_{3}{=}13.8$~hr, $P_{4}{=}10.4$~hr) mimic the behaviors of Waves 1 to 3 in the multi-sinusoidal model while the base-order component contributes little for recovering the observed modulations. The period of the base order far exceeds the rotation period of VHS 1256 b and is thus not physical. When we limit the base-order period below 25~hr, a conservative boundary set based on the 22.0~hr period measured in the Spitzer light curve, the truncated Fourier series does not provide a good fit to the observations.

\subsection{Lomb-Scargle Periodogram Analysis}

\begin{figure}[h]
  \centering
  \includegraphics[width=\columnwidth]{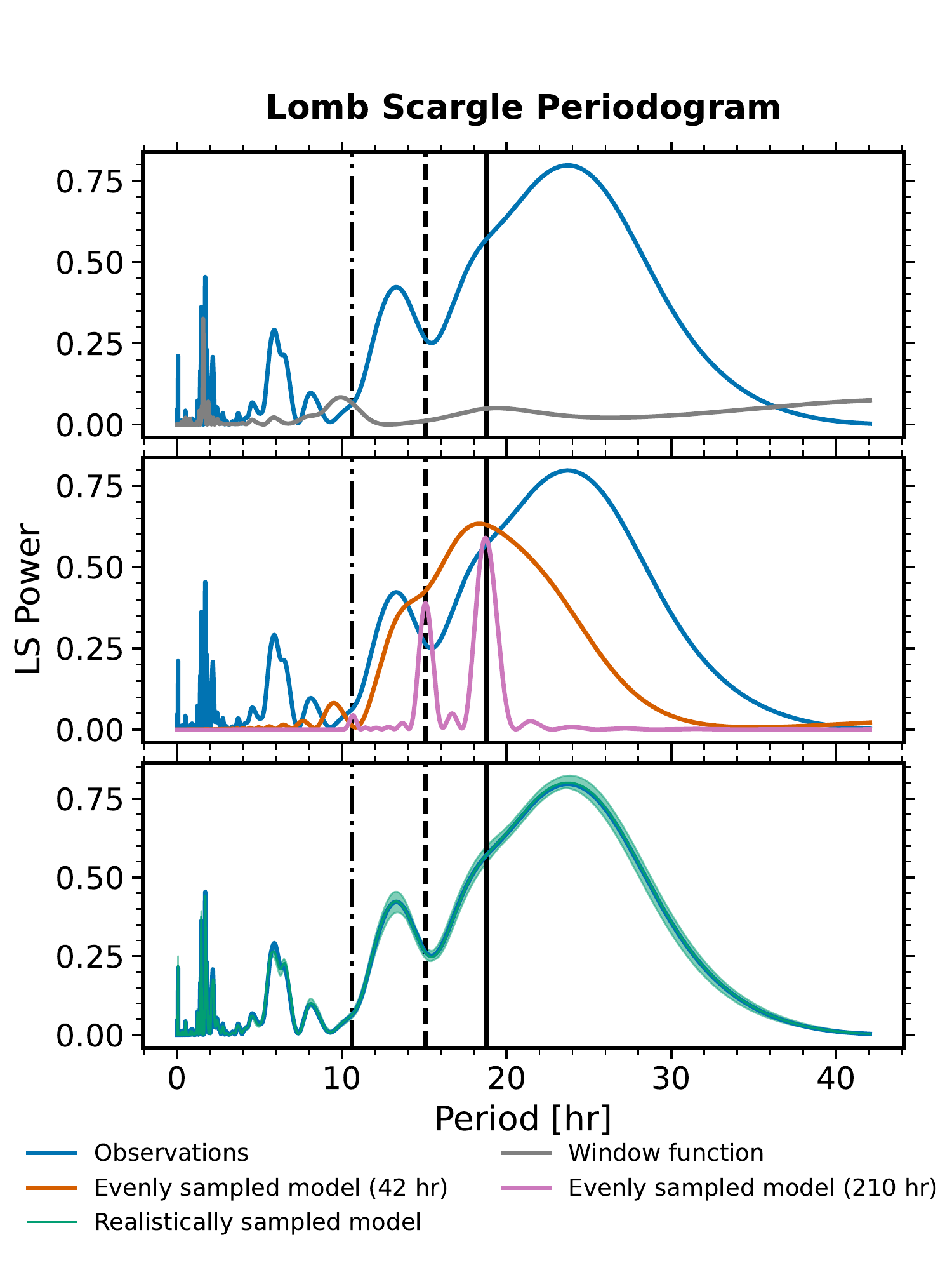}
  \caption{Results of the Lomb-Scargle periodograms analysis. The periodogram of the observed light curve (with the linear trend removed) is shown in all three panels as blue solid lines. In the upper panel, the gray solid line marks the periodogram of the observing window function. The middle panel includes two periodograms derived from the evenly sampled best-fitting model, with time ranges of \SI{42}{\hour} (orange) and \SI{210}{\hour} (magenta). In the bottom panel, the green shaded region shows the periodogram of the best-fitting model sampled in the same way as observations. The range of the region reflects fitting uncertainties. Due to the irregular sampling window, peak locations in the periodograms do not match the results obtained by fitting Equation~\ref{eq:3-2} (black vertical lines).}
  \label{fig:lomb-scargle}
\end{figure}

To further interpret the periodic signals, we conducted a Lomb-Scargle \citep{Lomb1976,Scargle1982a} periodogram analysis using the \texttt{LombScargle} module in the \texttt{astropy} package \citep{Robitaille2013} with the \texttt{fit\_mean} option turned on (see e.g., \citealt{Zechmeister2009} and \citealt{VanderPlas2018} and the references therein). The results are shown in Figure~\ref{fig:lomb-scargle}. Due to the irregular observing windows and a mixture of multiple periodic signals in \vhs's light curves, periods from fitting Equation~\ref{eq:3-2} do not match the peak positions in the periodogram. To understand this mismatch, we computed several illustrative periodograms and compare them with the one derived using the observed light curve. These additional periodograms are for:
\begin{enumerate}
  \item The window function: an evenly sampled Boolean series with 1 for visible windows and 0 for gaps.
  \item The over-sampled best-fitting model: a uniform grid over-sampled by $100\times$ and limited to a \SI{42}{\hour} window containing the observations.
  \item The over-sampled best-fitting model with a long baseline: a uniform grid over-sampled by $100\times$ and limited to a \SI{210}{\hour} window.
  \item The realistically sampled best-fitting model: sampled in the same manner and spanning the same time baseline as the observations.
\end{enumerate}

  These periodograms are presented in Figure~\ref{fig:lomb-scargle}. The upper panel shows the window function effect. By comparing the periodograms of the observed light curve and the window function, we find that HST's visibility cycle induces a series of peaks near \SI{96}{\minute} onto both periodograms. The middle panel shows that the limited observing time baseline combined with a mixture of multiple periodic signals can further confuse the periodogram. The same panel highlights the periodograms of two well-sampled best-fitting multiple-sinusoidal models with different lengths, \SI{42}{\hour} (the same as the observations) and \SI{210}{\hour} ($5\times$ the observation baseline). While the 210-hr periodogram exhibits the three peaks accurately, the 42-hr periodogram only detects the 18.6-hr periodic signal but fails for the other two shorter and weaker modulations. In the bottom panel, when the model light curve is sampled in the same manner as the observations, its periodogram regresses to exactly the same pattern as the observed one, which includes false positive peaks near 7 and 11 hr.

  The Lomb-Scargle periodogram of an irregularly sampled light curve that contains multiple periodic signals may contain false detections. As a result, systematic uncertainties in period measurement can significantly exceed the least-squares errors (see \S~\ref{sec:period-discussion}).

\subsection{Spectral Variability Analysis}
\label{sec:spec-fit}

\begin{figure*}
  \centering
  \includegraphics[width=\textwidth]{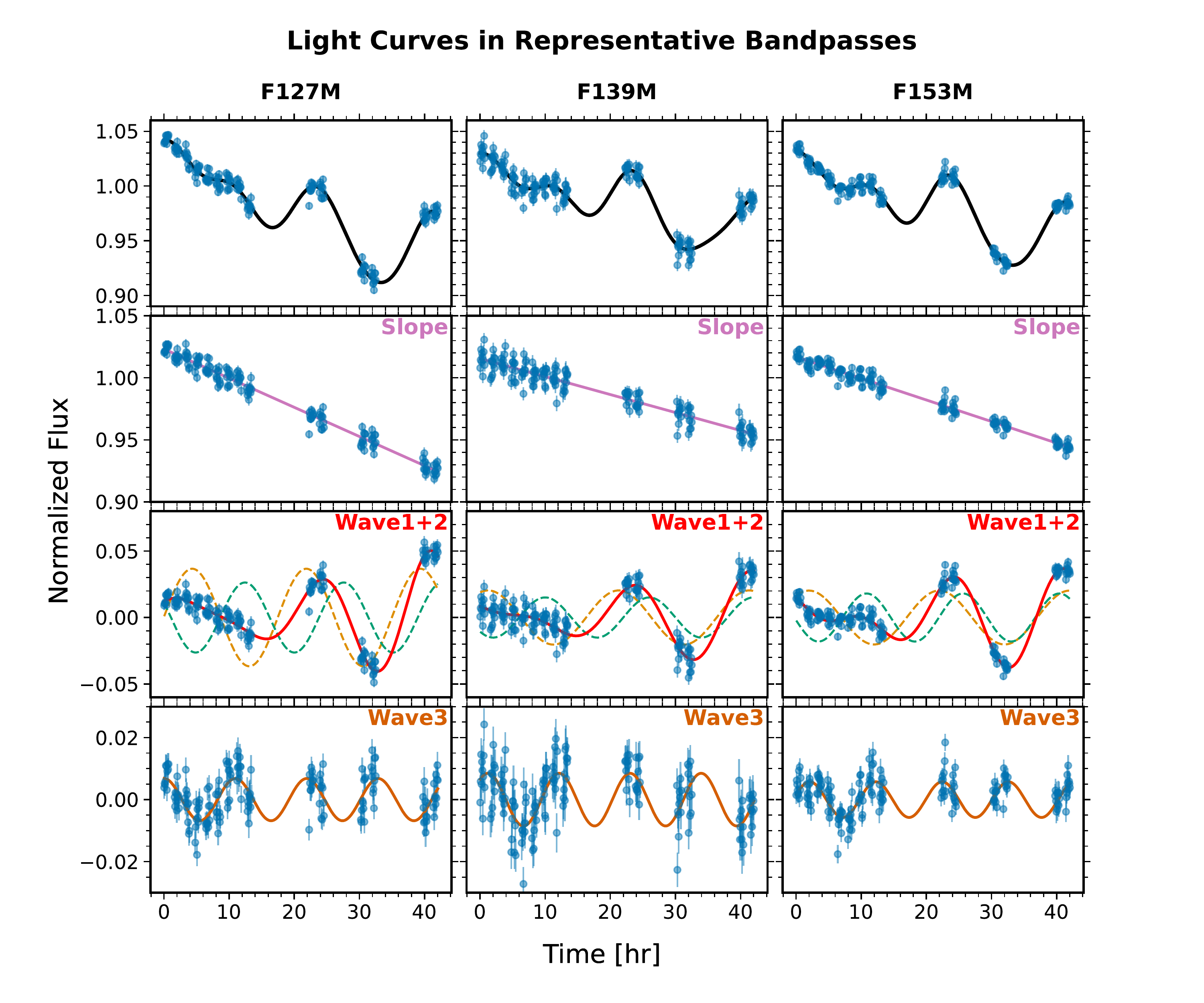}
  \caption{Fitting the multiple-sinusoidal model to light curves integrated in representative bandpasses. The three bandpasses represent the \SI{1.4}{\micro\meter} water absorption band (F139M, middle column) and the continuum (F127M, left column; F153M, right column). The markers are identical to those in Figure~\ref{fig:lcfit}. The top row shows the comparisons between model and observed light curves and Rows 2--4 illustrate individual components. In each row, the $y$-axis scale is kept consistent for all columns, allowing crude visual comparisons between light curves (e.g., the F127M light curve has steeper slope and high Waves 1 and 2 amplitudes; the F139M light curve has the highest Wave 3 amplitude, etc.)}
  \label{fig:planetary-wave-spectrum}
\end{figure*}

The wavelength-dependent variability is perceptible by a visual inspection. Between the maximum and minimum spectra (right panel of Figure~\ref{fig:specImage}), the relative flux difference decreases in the H$_{2}$O band around \SI{1.4}{\micro\meter}.  A comparison of light curves between the F127M, F139M, and F153M bands (Figure~\ref{fig:lightcurves}) reveals a more subtle spectral variation. In the first segment ($t < 12$~\si{\hour}), in which Waves 1 and 2 are in approximately opposite phases, the F127M light curve is nearly a straight line but the F139M and the F153M curves exhibit upward curvatures between $t=6$~\si{\hour} to \SI{12}{\hour}, suggesting that Wave 3 is more significant at longer wavelengths.

To quantify these findings, we fit the three-sine-wave model to the spectroscopically resolved light curves. These light curves are obtained from 30 wavelength bins evenly split the \SIrange{1.12}{1.65}{\micro\meter} range (bin size: $\Delta\lambda=$\SI{0.018}{\micro\meter}). In the spectroscopically resolved fits, the periods of the three sine waves are fixed to the values that best-fit to the broadband light curve. The amplitudes, phase offsets, and the linear slope are left as free parameters. The best-fitting parameters are determined in the identical manner as the broadband light curve fits.

\newcommand{\water}{\ensuremath{\mathrm{H_{2}O}}\xspace}

We find that the amplitudes of the three waves and the linear slope vary significantly with wavelength, but none of the waves show any chromatic changes in phase-offsets. Figure~\ref{fig:pw_model_comparison} shows the amplitudes as a function of wavelength for Waves 1 to 3 as well as the linear slope. Two patterns are revealed. The amplitudes of Waves 1 and 2 and the slope of the linear trend decrease in the \SI{1.4}{\micro\meter} \water band. This spectral variability pattern has been identified in a handful of L/T transition dwarfs \citep[e.g.,][]{Apai2013,Yang2014,Zhou2018}, in red late-L type dwarfs \citep[e.g.,][]{Lew2020}, and in \vhs in its Epoch 2018 observations \citep{Bowler2020,Zhou2020a}. The amplitude reduction in the \water band has been reproduced by partly cloudy models where the optical depths of the Fe and Si clouds are greater in the continuum than in the \water absorption band \citep{Morley2014}.


The amplitude of Wave 3 shows an opposite wavelength-dependence and increases in the \water band. Based on atmospheric models, this signal may indicate the presence of thermal profile anomalies \citep{Morley2014,Robinson2014}. In these models, the anomalous thermal profile ($T$-$P$)  contains a constant displacement above a pressure threshold. This local $T$-$P$ variation affects the strength of molecular absorption more than the continuum and create greater modulations in the absorption bands. Thermal profile anomalies were put forward to explain the spectral modulations of the T6 dwarf 2MASS~J22282889-4310262 in which the continuum and \water bands modulate in opposite phases  \citep{Buenzli2012,Robinson2014}.

\subsection{Comparing the Observed Spectral Variability with Models}

\begin{figure*}
  \centering
  \includegraphics[width=0.48\textwidth]{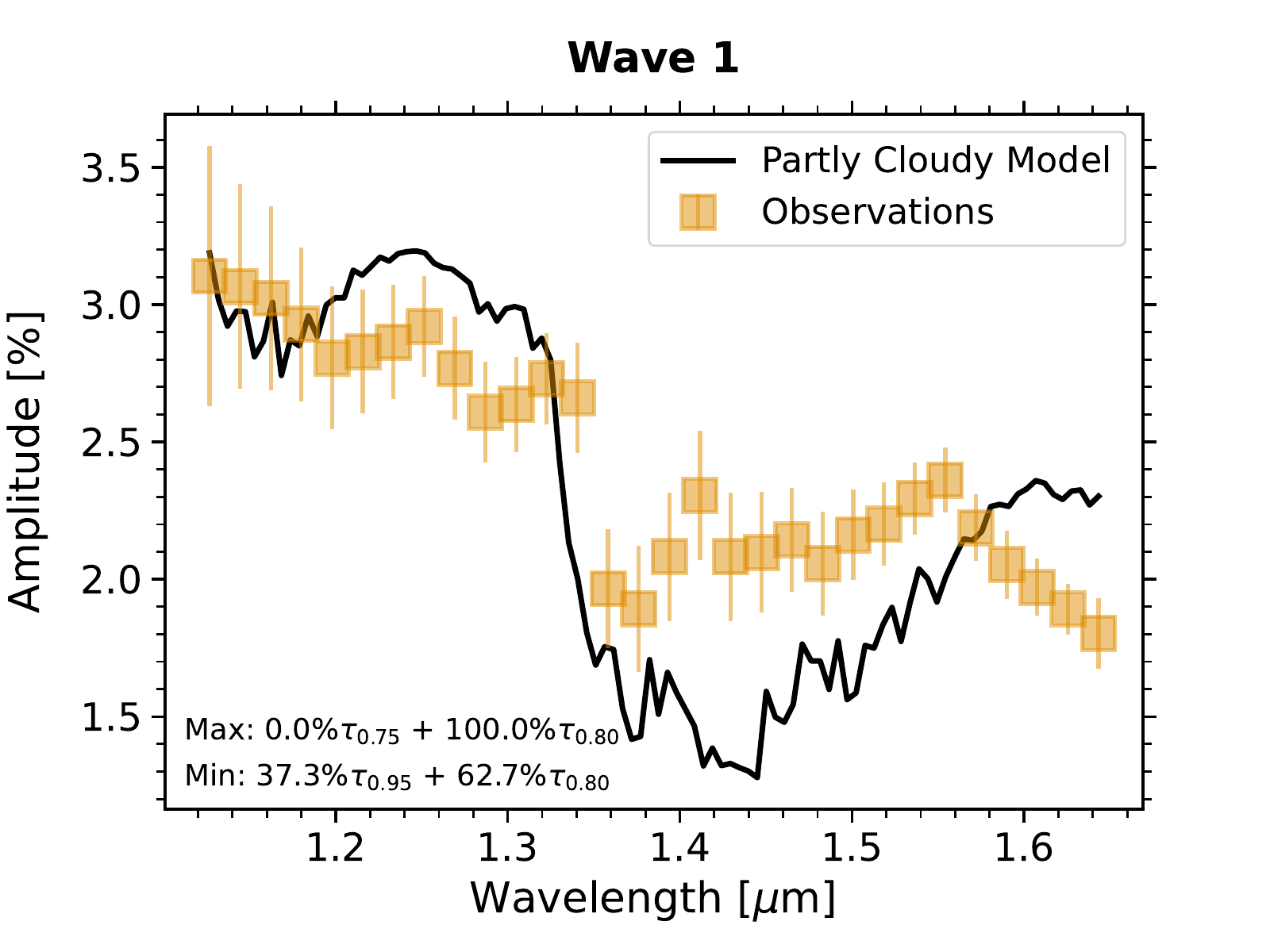}
  \includegraphics[width=0.48\textwidth]{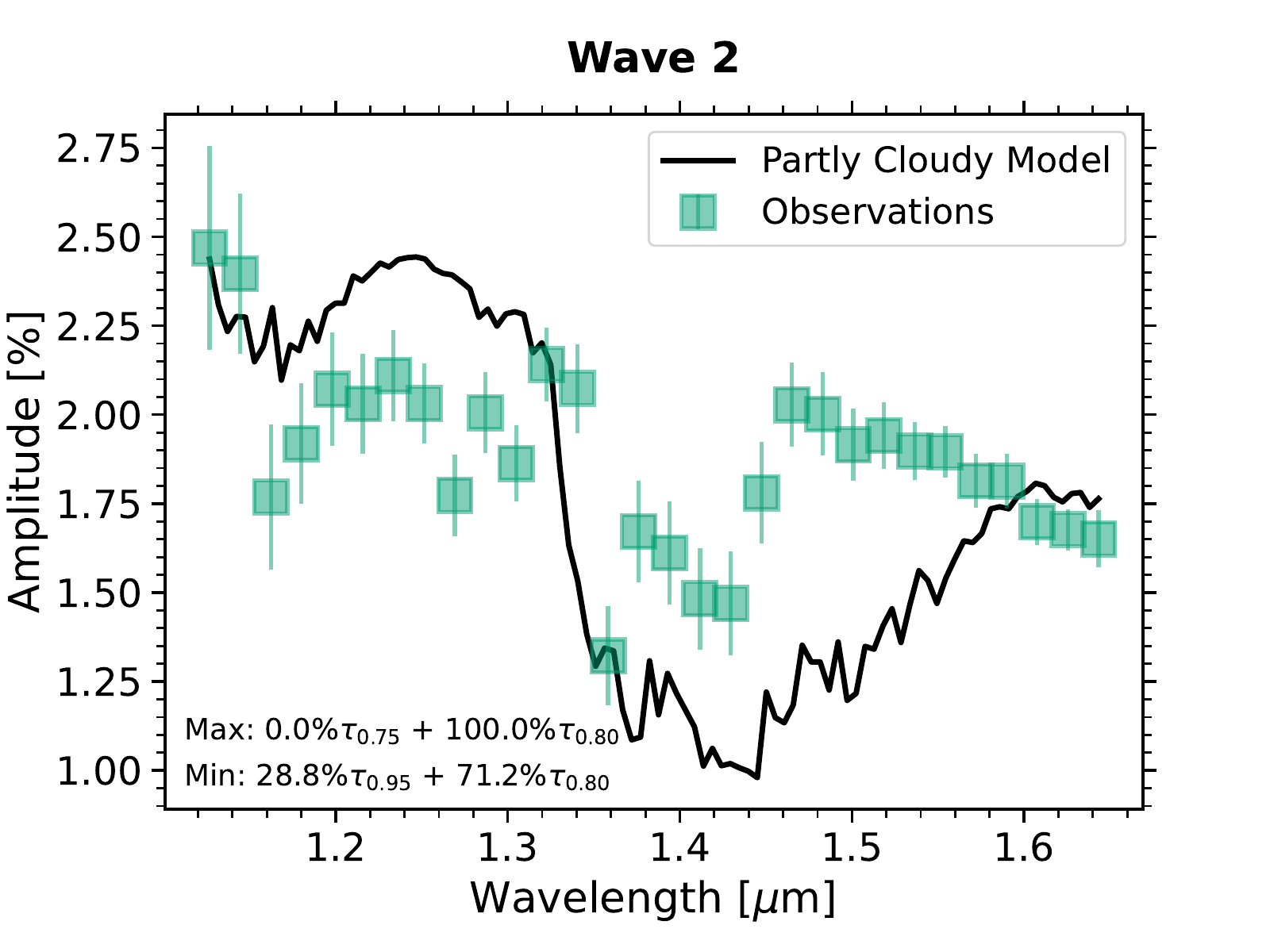}
  \includegraphics[width=0.48\textwidth]{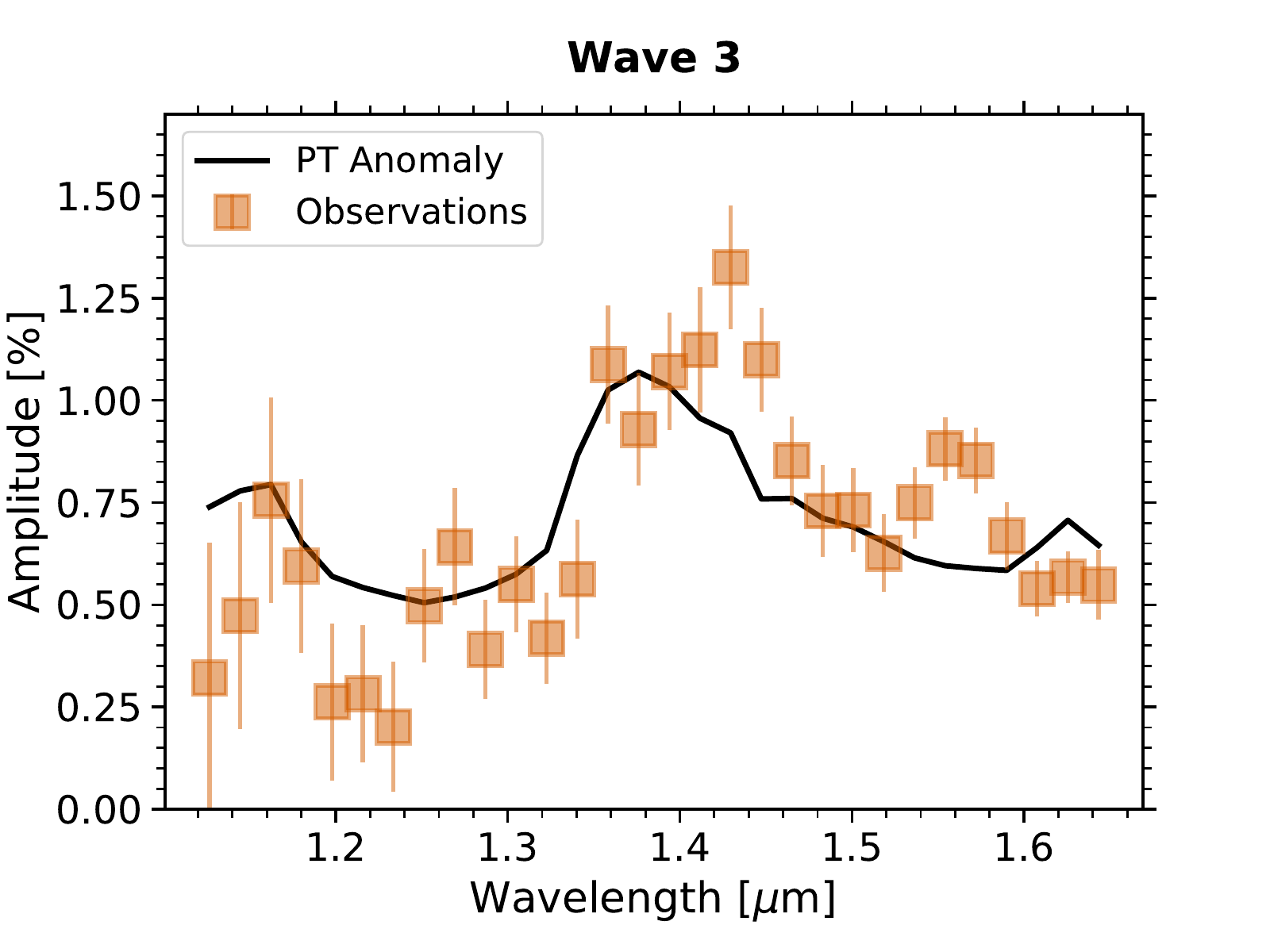}
  \includegraphics[width=0.48\textwidth]{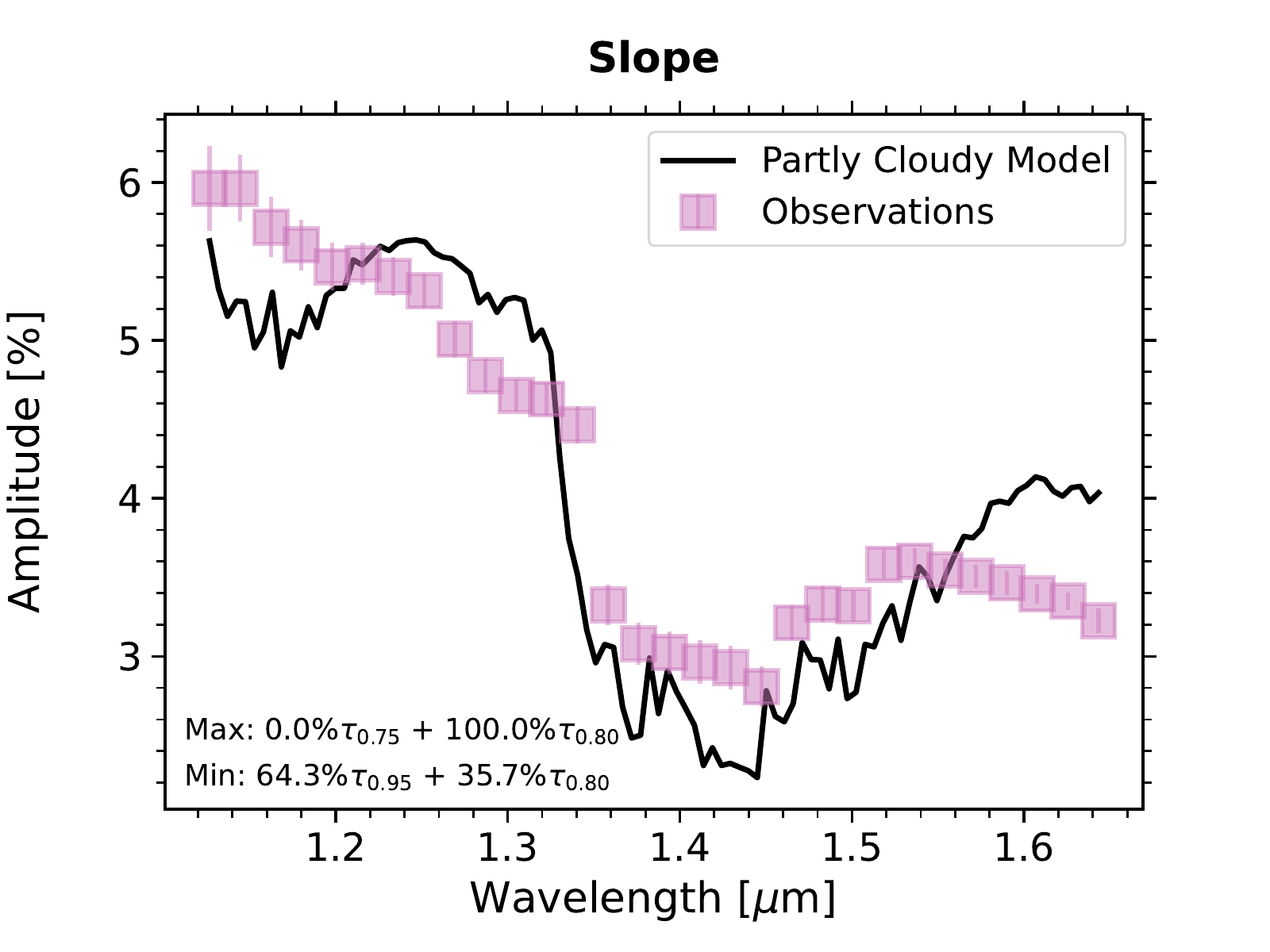}
  \caption{Comparisons of the observed wavelength-dependent amplitudes and atmospheric model predictions. The spectral variability patterns of the sine waves fall into two categories: Wave 1, Wave 2, and the linear trend match well with the prediction of a partly cloudy model, while Wave 3 is better fit with a model involving thermal profile anomalies.}
  \label{fig:pw_model_comparison}
\end{figure*}
\newcommand{\tauc}{\ensuremath{\tau_{\mathrm{c}}}}
In \citet{Bowler2020} and \citet{Zhou2020a}, the partly cloudy model successfully reproduced \vhs's spectral variability observed in the Epoch 2018 data. This model consists of two hemispheres with different cloud optical depths relative to the fully cloudy model (\tauc). The less cloudy hemisphere has a \tauc of 75\% of the fully cloudy optical depth and the more cloudy hemisphere has a \tauc of 90\%. All other parameters are identical to those best-fit to the mean spectrum ($\teff=$ \SI{1000}{\kelvin}, $\log g = 3.2$, $f_{\mathrm{sed}}=1.0$, and $\tauc=80\%$). Here, we compare this model with the spectral variability curves of Waves 1, 2, and the linear slope.

To allow the overall amplitudes to vary to fit the observations, we introduce covering fractions of the less cloudy ($\tauc=75\%$) and the more cloudy ($\tauc=90\%$) patches as free parameters. Then, the model's maximum and minimum spectra are:
\begin{align}
  \label{eq:7}
  &S_{\mathrm{max}} = f_{1}S_{\tauc 75} + (1 - f_{1}) S_{\tauc 80},\\
  &S_{\mathrm{min}} = f_{2}S_{\tauc 90} + (1 - f_{2}) S_{\tauc 80},
\end{align}
where $S_{\tauc 75}$, $S_{\tauc 80}$, and $S_{\tauc 90}$, represent model spectra for $\tauc=75\%$, 80\%, and 90\%, respectively; $f_{1}$ and $f_{2}$ are the covering fractions of the  $\tauc=75\%$ and 90\% patches. The predicted semi-amplitude is
\begin{equation}
  \label{eq:8}
  \Delta F/F = (S_{\mathrm{max}} - S_{\mathrm{min}}) / (S_{\mathrm{max}} + S_{\mathrm{min}}).
\end{equation}

We fit the output of Equation~\ref{eq:8} to the observed spectral variability curves of Waves 1, 2, and the linear slope and show the results in Figure~\ref{fig:pw_model_comparison}. The model reproduces the overall trends, including the reduced amplitudes in the water band and the decrease in amplitudes at longer wavelengths, but does not fit the observed curves perfectly. In particular, the partly cloudy model over-predicts the difference between the water band and the continuum. This result confirms that Waves 1, 2 and the slope trace the change of cloud optical depths. However, our untuned atmospheric model does not completely characterize the clouds of \vhs.

We also compare a \citet{Morley2014} hot spot model to the spectral variability curve of Wave 3. In this model, extra energy is injected at $P=10$~\si{bar} with a Chapman heating function to elevate the thermal profile above this pressure level in one hemisphere. In the other hemisphere, the thermal profile is kept in the form determined by radiative-convective equilibrium. The spectra of the two hemispheres and the corresponding spectral variability curve are derived using Equations~\ref{eq:7} and \ref{eq:8}. We allow the model $\Delta F/F$ curve to scale linearly to fit the observed results. The scaling corresponds to the covering fraction change of the spot. As shown in the lower left panel of Figure~\ref{fig:pw_model_comparison}, the model matches the observed curve well, supporting that Wave 3 probes a heterogeneous distribution of thermal profiles.

\section{The Long-term Light Curve Evolution}
\label{sec:long-term}
\begin{figure*}[th]
  \centering
  \includegraphics[width=\textwidth]{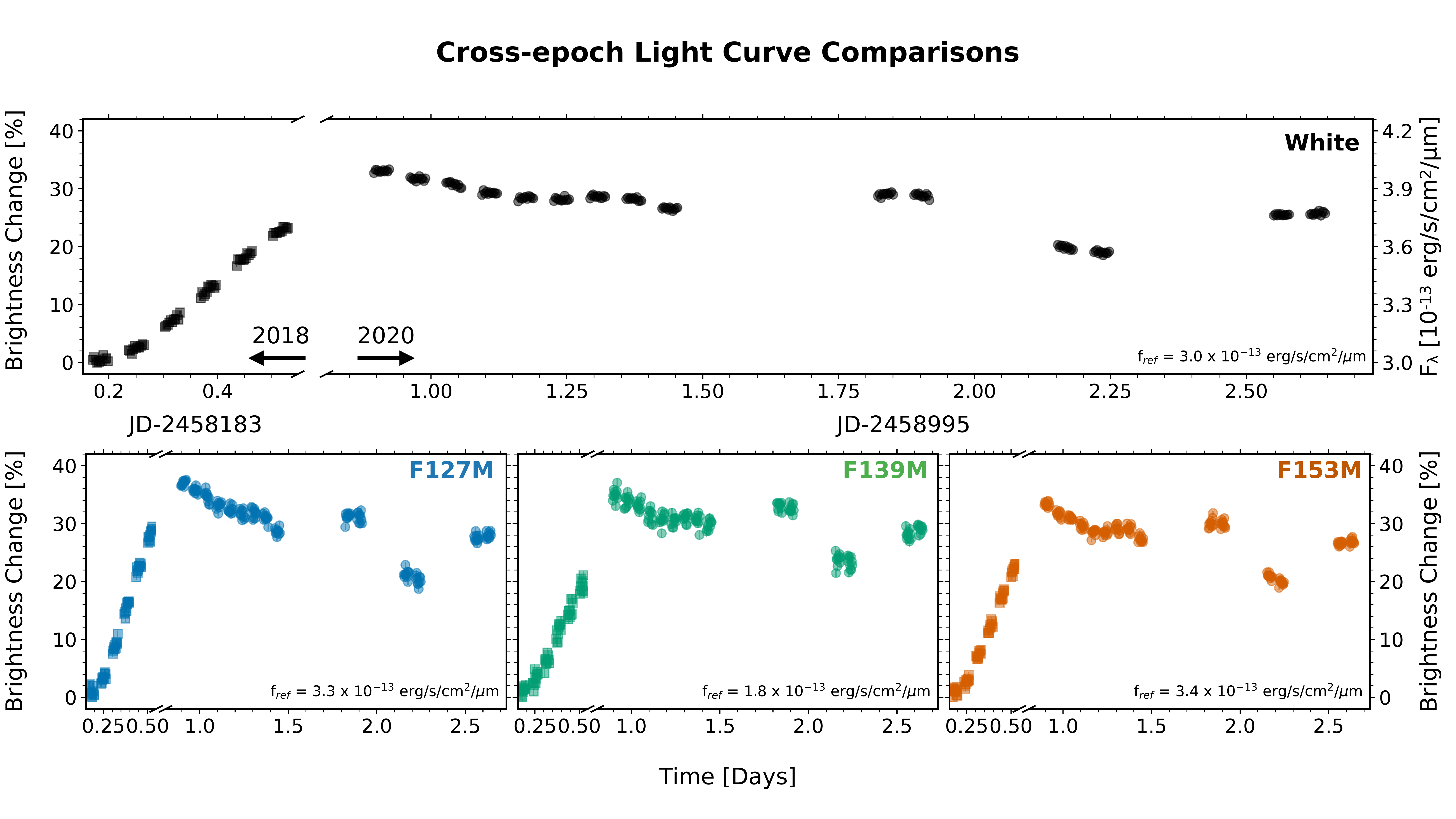}
  \caption{Combined light curves of \vhs in the 2018 and 2020 epochs. In four representative bands: G141 broadband (upper panel), F127M (centered at \SI{1.27}{\micro\meter}, sampling the continuum on the blue side of the water band; lower-left panel), F139M (centered at \SI{1.39}{\micro\meter}, sampling the water band, lower-middle panel), and F153M (centered at \SI{1.53}{\micro\meter}, sampling the continuum on the red side of the water band; lower-right panel). The Epoch 2018 light curves are plotted as squares and the Epoch 2020 light curves are shown as circles. The light curves are normalized by the minimum flux measured in each band. The reference flux values are indicated at the bottom right corners. The right axis of the upper panel marks the broadband flux density in erg/s/cm$\mathrm{^{2}}$/$\mu$m. The $x$-axes are broken to fit the two epochs of data.}
  \label{fig:longtermlc}
\end{figure*}

\begin{figure*}[th]
  \centering
  \includegraphics[width=0.48\textwidth]{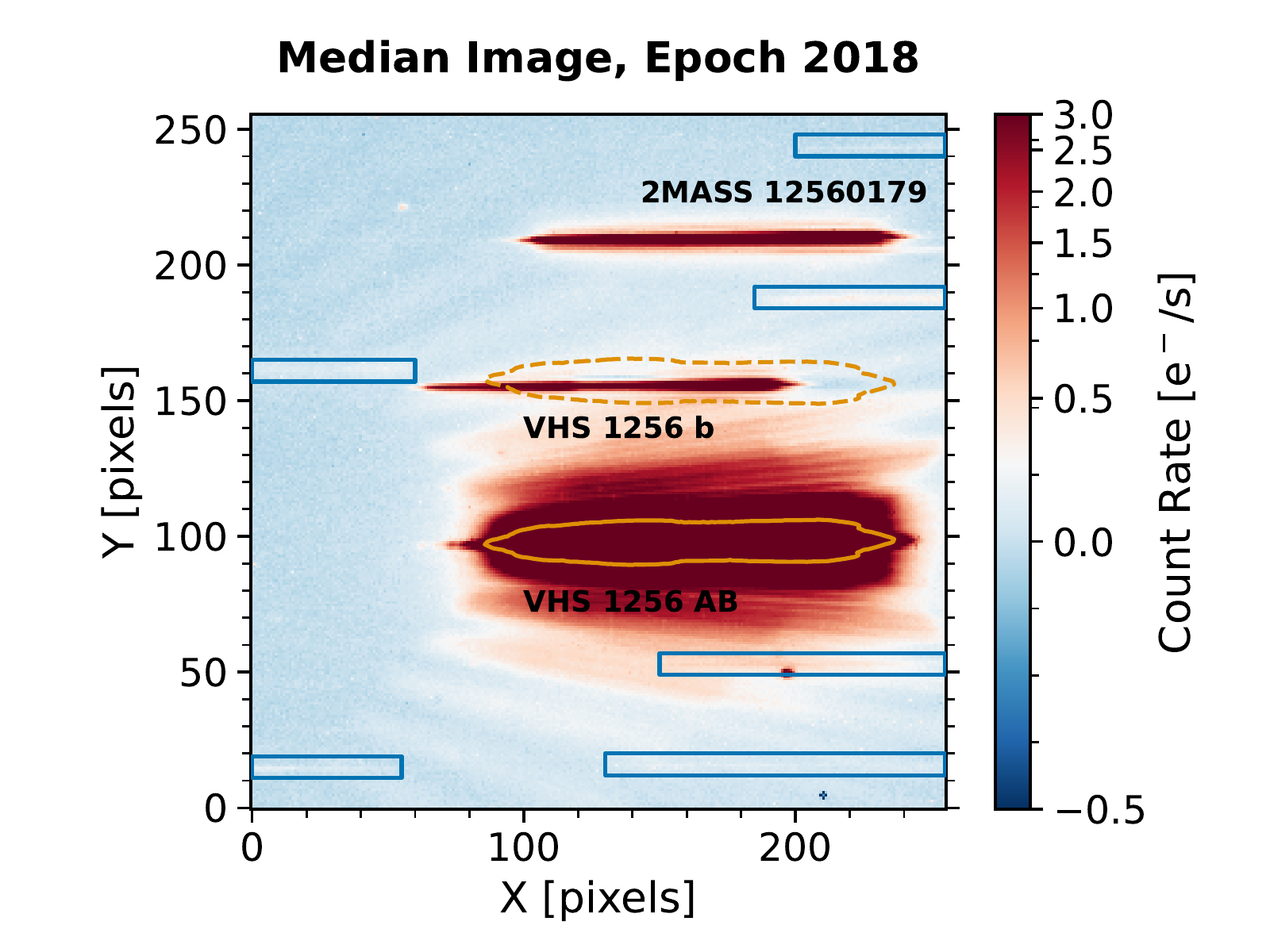}  \includegraphics[width=0.48\textwidth]{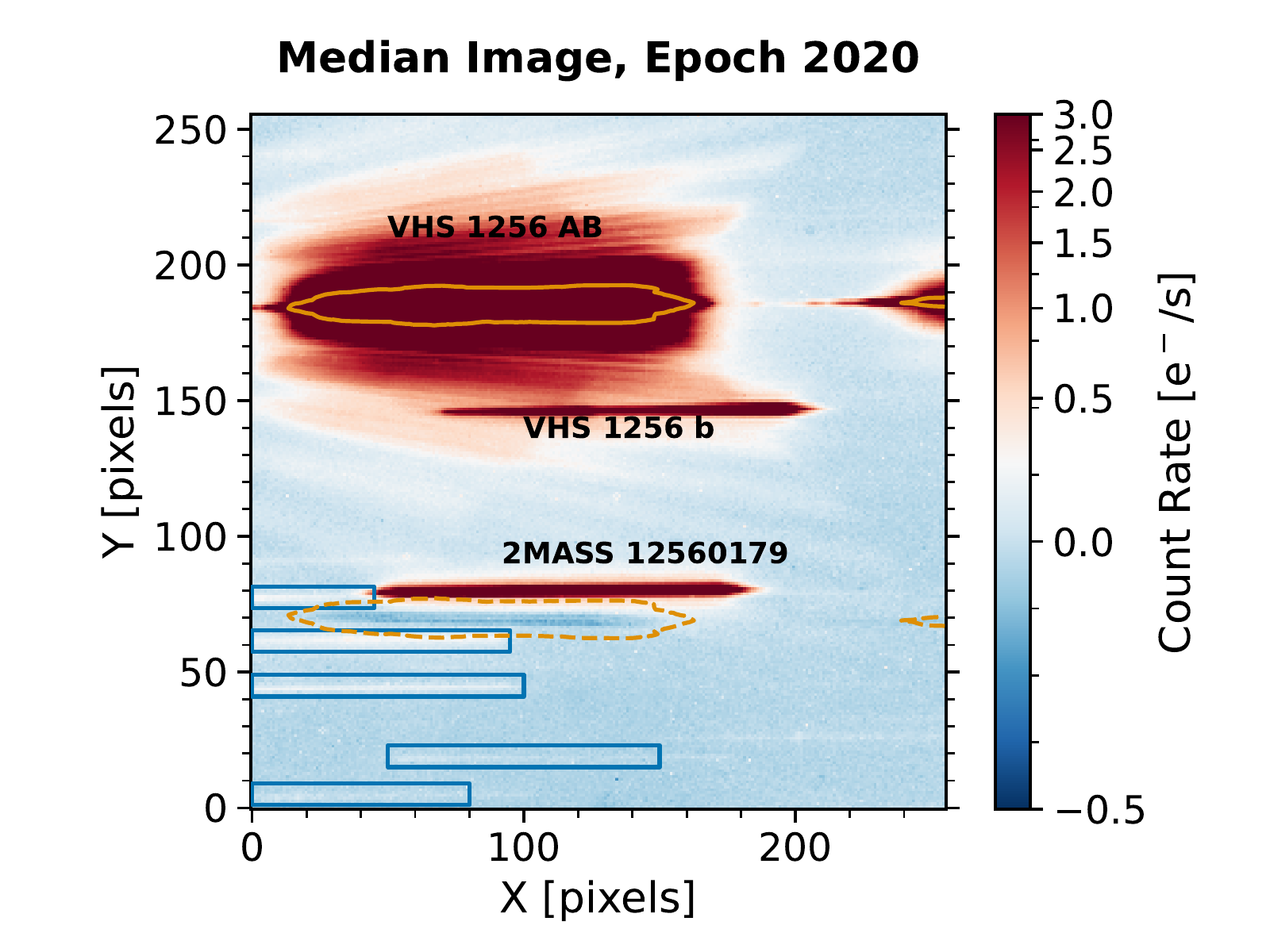}
  \caption{Heavily stretched grism images of \vhs showing the contaminating sources in the Epoch 2018 (left) and Epoch 2020 (right) observations. The orange solid contours mark a flux level at 1\% of the image maximum, highlighting the position of VHS~1256~AB. The primary binary's extended PSF is the major contaminating source in the flux measurements of \vhs. The orange dashed contours show the position of the cross-talk systematics caused by the intense illumination from VHS~1256~AB. The cross-talk signal is at a symmetric location relative to the PSF of VHS~1256~AB with respect to the $x$-axis and has a negative flux. The blue rectangles mark the positions of spectral traces of faint background sources. }
  \label{fig:contamination}
\end{figure*}

The two epochs of HST observations of \vhs enable an investigation of its spectral variability on the \SIrange{1}{2}{yr} timescale. In this Section, we conduct a joint analysis of the two spectral time series datasets. This includes: 1) consistently reducing the two epochs of data; 2) evaluating cross-epoch flux differences; 3) estimating errors in cross-epoch flux calibration; and 4) interpreting the spectral variability of \vhs between the two epochs.

\subsection{Consistent Data Reduction}

Although the two epochs of observations were conducted consistently in almost every respect, the difference in the telescope position angle can introduce subtle systematic errors in the cross-epoch flux calibration. These errors come from three sources: contamination from the primary VHS~1256~AB; contamination from background sources; and detector systematics. Among these, the difference in the primary star's contaminating flux is the most significant error source and can be mitigated by subtracting the primary PSF consistently.

In \citet{Bowler2020}, the light curves of \vhs were extracted without primary subtraction. This decision was made  based on the fact that the contaminating flux has a similar intensity as the sky background and primary subtraction did not significantly affect the \emph{relative} variability measurements. However, cross-epoch calibration requires \emph{absolute} photometry and thus primary subtraction becomes necessary. In the Epoch 2018 observations, the primary binary's PSF contributes 7.8\% of the flux relative to the average spectrum of \vhs in a $R=4$~pixels extraction window. Without primary subtraction, the contamination would cause biases in measuring the absolute flux. Therefore, instead of directly adopting the results of \citet{Bowler2020}, we re-reduced the Epoch 2018 data with the ``flip-and-subtraction''  method (described in \S \ref{sec:obs}) and extracted \vhs's  spectra from the primary-subtracted spectral images.

 Figure~\ref{fig:longtermlc} shows the consistently reduced light curves of Epoch 2018 and Epoch 2020. These light curves are not phase-folded, because the current rotational period constraint does not permit precise phase propagation for nearly 900 rotation periods\footnote{Linearly scaling the uncertainty of the Spitzer light curve period by 900 yields \SI{20}{\hour}, the same order as the rotation period itself.}. The flux changes are measured relative to the lowest flux point, which occurred at the beginning of Epoch 2018.

\vhs is brighter in Epoch 2020 than Epoch 2018 by $16.9\pm1.0$\% in time-averaged G141 broad band flux. Relative to the median flux, the broad band peak-to-peak flux difference is $33.3\pm0.2$\%. In the three medium filter bands, the peak-to-peak flux differences decrease slightly with wavelength: $37.6\pm0.5$\% in F127M, $37.1\pm1.0$\% in F139M, and $33.9\pm0.5$\% in F153M. We note that the quoted errors only include those directly derived from photometry: photon noise, sky background, dark current, and flat field uncertainties. In the next subsection, we discuss additional systematic uncertainties in cross-epoch flux calibration in the next subsection.

\subsection{Systematic Uncertainties in Cross-Epoch Flux Calibration}

Apart from the error due to primary contamination, there are two additional sources of systematic uncertainty that can interfere with cross-epoch flux calibration. The first one is contamination from unknown background sources. As shown in Figure~\ref{fig:contamination}, several faint spectral traces are present in the median-combined images. Because of the telescope PA difference between the 2018 and 2020 observations, one faint background spectrum may contaminate \vhs in one epoch but not the other.

To estimate the contribution of faint contaminating flux to the flux calibration of \vhs, we visually selected background spectral traces (marked by blue rectangles in Figure~\ref{fig:contamination}) and calculated their average contaminating flux in the extraction window ($R=4$~pixels). Among all sources, the average flux is  $0.31\mathrm{e}^{-}\mathrm{s^{-1}/column}$, corresponding  to 1.4\% of \vhs's average flux in the G141 broadband or 2.2\% of the average flux in the water absorption band. These values are likely upper bounds, because no background sources are identified near or overlapping with \vhs. Only in the unlucky scenario where one background source occupies the exact same pixels as \vhs can contamination reach these levels.

The second uncertainty source is detector systematics. The flat field uncertainty of WFC3/IR is about 1\% per pixel. In the broadband measurements that involve more than 400 pixels, the flat field errors are attenuated to below 0.1\% levels, and thus are unlikely to affect our results. However, one rarely discussed WFC3/IR systematic, cross-talk, may have a more significant effect in cross-epoch flux calibration. Cross-talk is manifested as a low-level negative and mirrored image of a bright source on the detector. The negative image is at a mirrored position with respect to the $x$-axis relative to the positive image. In the Epoch 2018 observations, the cross-talk image of VHS 1256 AB happened to partially overlap with the spectral trace of \vhs, possibly reducing the observed flux. To our knowledge, there is no available software to correct for this effect.

To estimate the uncertainty due to cross-talk, we locate the negative image in the Epoch 2020 data by mirroring the coordinates of the positive image and calculate the contaminating flux in the extraction window. The average cross-talk count rate is $-0.37,\mathrm{e^{-}/s/column}$, corresponding to 1.6\% of \vhs's average G141 broadband flux or 2.5\% of its average water band flux.

Combining the two sources of uncertainty in quadrature, we find the systematic uncertainty in cross-epoch flux calibration is $-0.49\,\mathrm{e^{-}/s/column}$, corresponding to 2.2\% of \vhs's average G141 broadband flux or 3.3\% of its average water band flux. The systematic uncertainty is therefore significantly greater than the photometric uncertainty. The uncertainty estimates corroborate what we observed with the background star 2MASS J12560179$-$1257390, which shares a similar brightness as \vhs. The background star's light curves are flat in both epochs, but the epoch-averaged flux differ by 1.5\% between the two. Based on this analysis, we revise the uncertainties in the cross-epoch flux comparison: $33.3\pm2.2$\% in G141 broad band, $37.6\pm2.2$\% in F127M, $37.1\pm3.4$\% in F139M, and $33.9\pm1.7$\% in F153M.

\subsection{Cross-epoch Spectral Variability}

\begin{figure*}[t]
  \centering
  \includegraphics[width=0.48\textwidth]{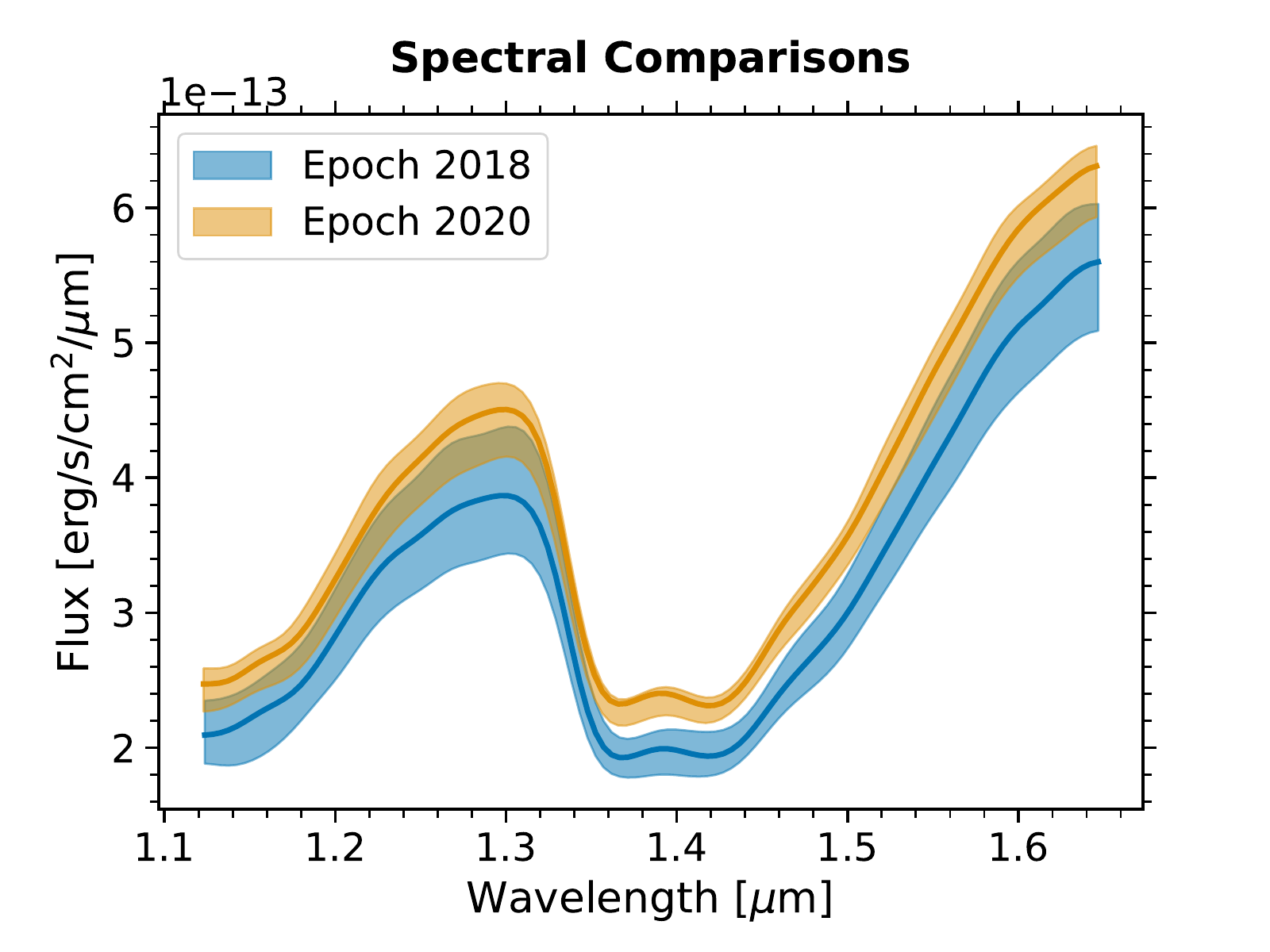}
  \includegraphics[width=0.48\textwidth]{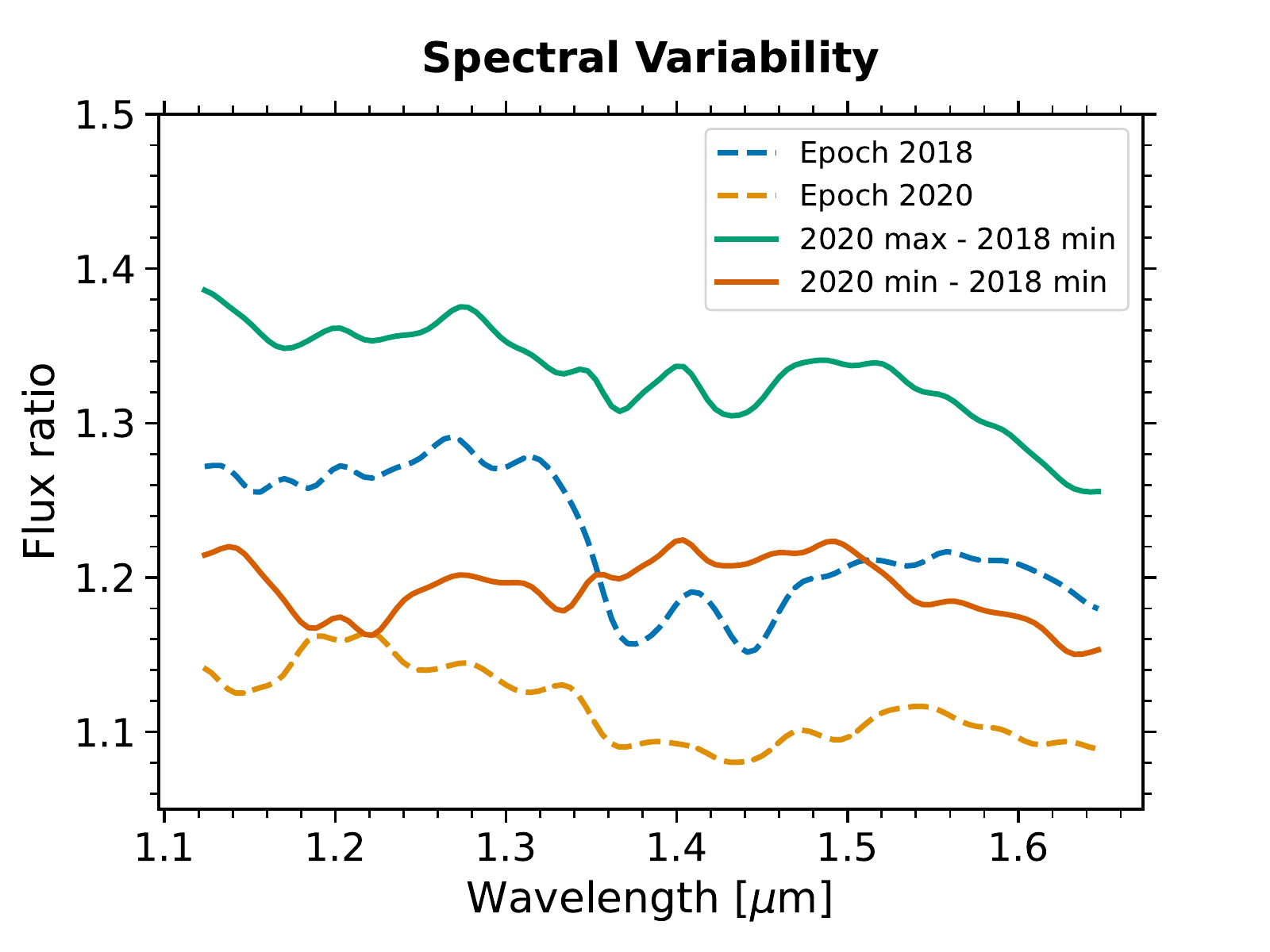}
  \caption{Wavelength-dependent flux variations between Epoch 2018 and Epoch 2020. Left: the spectra of \vhs in Epoch 2018 (blue) and Epoch 2020 (yellow). The solid lines are for median spectra and the dashed lines are for extrema spectra. The shaded regions show the range of spectral variability. Right: representative spectral variability curves. The solid lines show the $F_{\mathrm{max}}/ F_{\mathrm{min}}$ flux ratios between the Epoch 2020 max and the Epoch 2018 min (green), as well as between the Epoch 2020 min and the Epoch 2018 min (red). For reference, flux ratio curves derived using the extrema of the same epochs are shown in dashed lines (blue for Epoch 2018; yellow for Epoch 2020).}
  \label{fig:longtermspec}
\end{figure*}

To further examine cross-epoch spectral variability, we derive the flux ratio between the two epochs (Figure~\ref{fig:longtermspec}). Two cross-epoch comparisons between extrema are presented: Epoch 2020 maximum vs. the Epoch 2018 minimum; and Epoch 2020 minimum vs. Epoch 2018 minimum. For reference, two same-epoch spectral variability curves are over-plotted in Figure~\ref{fig:longtermspec}. As discussed in \citet{Bowler2020} and \citet{Zhou2020a} and in previous sections of this paper, both of the two same-epoch curves have lower variability in the \SI{1.40}{\micro\meter} \water band than continuum, supporting that heterogeneous clouds are the main driver for the spectral variability. However, the cross-epoch spectral variability curves differ from the same-epoch curves. The 2020 max vs. 2018 min flux ratio curve has a gradual trend that decreases from 1.38 at \SI{1.1}{\micro\meter} to 1.25 at \SI{1.65}{\micro\meter} without exhibiting a significant amplitude decrease in the \water band.
Spectral variability with weak wavelength dependence in the WFC3/G141 bandpass was observed  brown dwarfs \citep[e.g.,][]{Yang2014,Manjavacas2017,Lew2020b}, as well as the planetary mass object PSO~J318.22 that has an identical spectral type as \vhs \citep{Biller2017}. Models invoking heterogeneous haze layers or modified local temperature gradients have been proposed to reproduce this spectral change \citep{Yang2014,Tremblin2020}.

\section{Discussion}
\label{sec:discussion}

\subsection{Interpreting the Periodicity of \vhs's Light Curves }
\label{sec:period-discussion}

\begin{figure}[ht]
  \centering
  \includegraphics[width=\columnwidth]{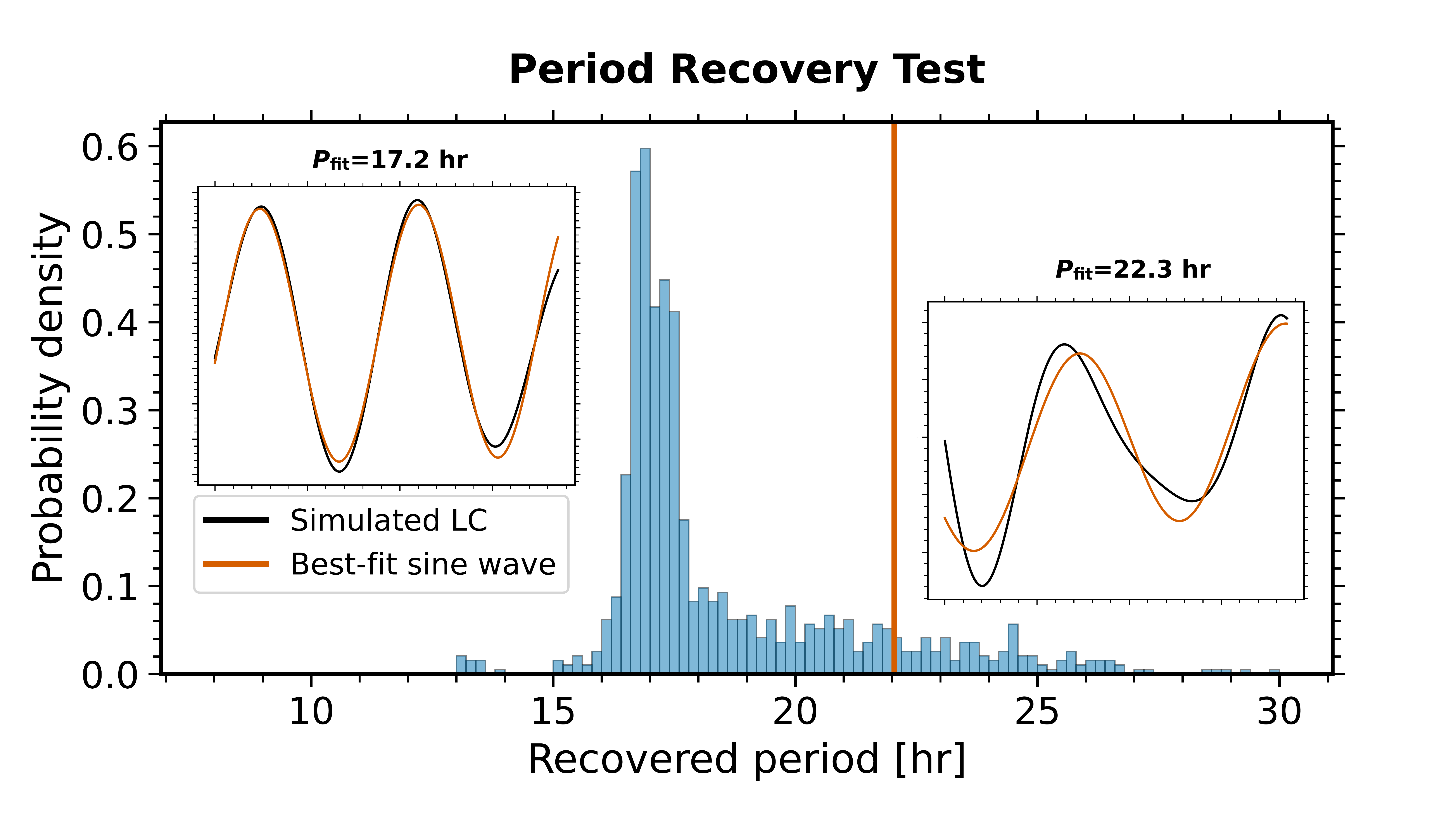}
  \caption{The distribution of the best-fitting periods of a 36~hr light curve segment obtained from a period-recovery test.  In this test, a single sine wave is fit to 10,000 random 36-hr segments of \vhs's evolving light curve, which is assumed to be the multiple-sinusoidal model that we found in the Epoch 2020 data. We show the results with a histogram that is normalized so that an integral of the probability density equals to 1. The distribution peaks at 17~hr, the average period of Waves 1 and 2. It also has a long tails extending towards long periods, suggesting a non-negligible probability to detect a signal with period $P_{\mathrm{fit}}\ge22$~\si{hr}. Two embedded panels show examples of light curve segments fit by models with different periods. Based on this result, we can at least partially attribute the apparent discrepancy in period measurements between the HST and Spitzer observations to the limited observing windows.}
  \label{fig:period_recovery_test}
\end{figure}

\vhs's light curves exhibit multiple periods that differ from each other significantly. These periods include: $22.04\pm0.05$\,\si{hour} (Spitzer \SI{4.5}{\micro\meter}, \citealt{Zhou2020a}),  $18.8\pm0.2$\,\si{\hour} (Wave 1), $15.1\pm0.2$\,\si{\hour} (Wave 2), and $10.6\pm0.1$\,\si{\hour} (Wave 3). Additionally, fitting a single sinusoid to the 9-hr long Epoch 2018 light curve yields a period of $22.5\pm0.5$ hr, similar to the one determined by the Spitzer observations.  Periodicity in brown dwarf light curves are often interpreted as rotation rates \citep[e.g.,][]{Metchev2015,Zhou2016, Zhou2020a}. The apparently incompatible measurements demonstrate the challenges in precisely determining the rotation periods using light curves.  We discuss the origins of discrepancies between these periods and the implications on rotation measurements.

The circulation patterns probed by a finite-length light curve are incomplete. When estimating rotation periods with light curve fitting, unaccounted model complexity can lead to biased results. For example, if the true light curve during the Spitzer observation is our best-fitting three waves but modeled (inaccurately) by a single sinusoid, the probability of recovering a period of $P\ge22$~\si{\hour} is considerable (Figure \ref{fig:period_recovery_test}). Likewise, substantial systematic uncertainties can occur in fitting the HST light curves with the multiple-sinusoidal models. If we replace the linear trend, which is essentially a first-order approximation of an unknown long-term variation, with a long-period ($42\,\mbox{hr}< P< 210\,\mbox{hr}$) sinusoid, the best-fitting periods of the three waves can vary by ${\sim}$1~hr (5 to 10\% relative errors), driving the period of Wave 1 closer to the Spitzer period in several cases. These results show that the systematic uncertainties caused by fitting an incomplete model can considerably exceed the least-squares-determined uncertainty when insufficient monitoring does not reveal the light curve's underlying evolving patterns.

\begin{figure}[t]
  \centering
  \includegraphics[width=\columnwidth]{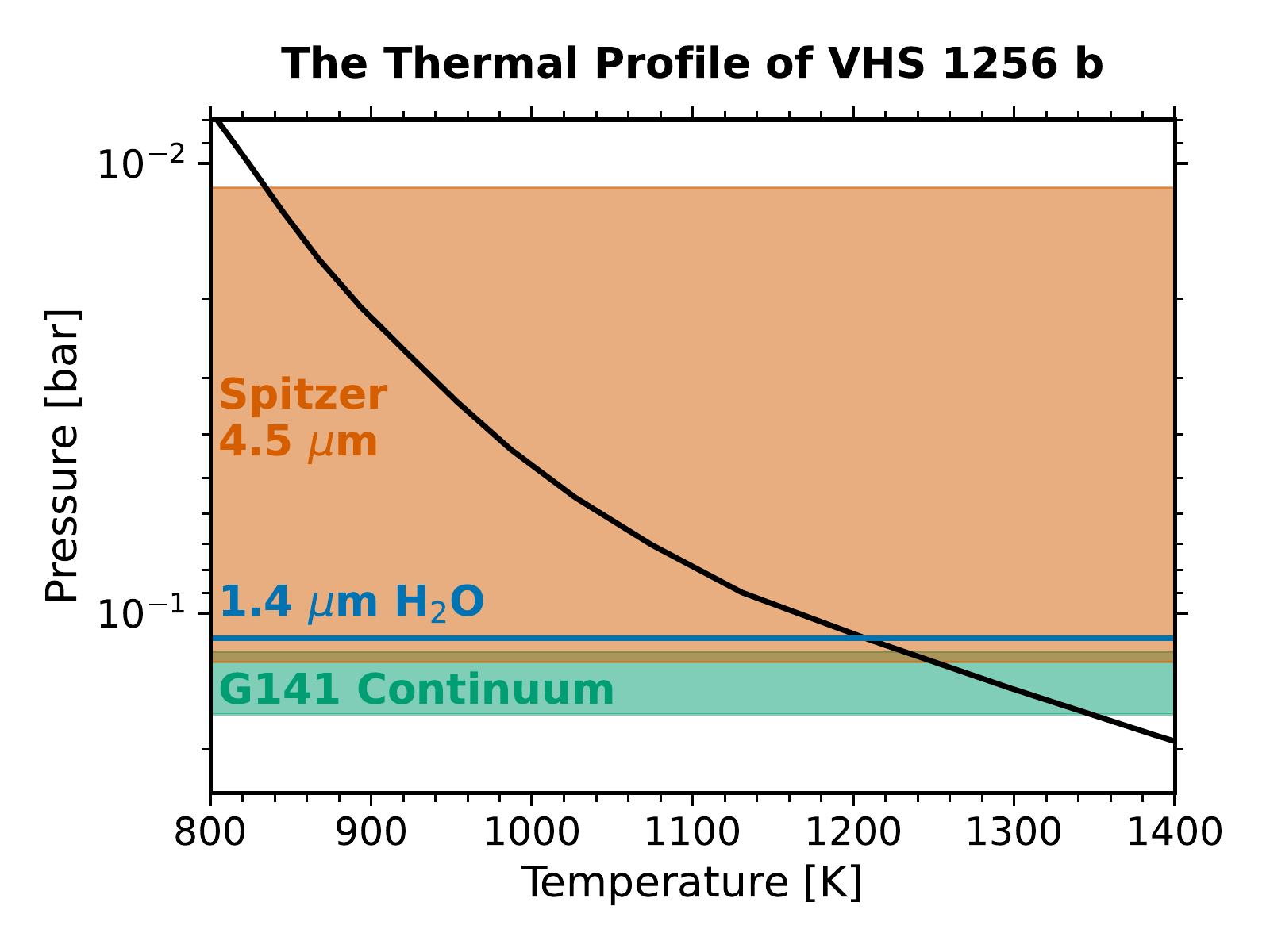}
  \caption{The thermal profile of \vhs and the pressure levels probed by various bandpasses. The black solid line is the $T\mbox{-}P$ profile of the best-fitting atmospheric model found by \citet{Zhou2020a} ($T_{\mathrm{eff}}=1000$~\si{\kelvin}, $\log g=3.2$, $f_{\mathrm{sed}}=1.0$, 80\% cloud optical depth of the fully cloudy model). The green, blue, and orange horizontal bands/lines mark the pressure levels probed by WFC3/G141 continuum ($\lambda < 1.30$~\si{\micro\meter} or $\lambda > 1.50$~\si{\micro\meter}), the \SI{1.4}{\micro\meter} \water{} band, and the Spitzer \SI{4.5}{\micro\meter} channel, respectively. These pressure levels are mapped from the observed brightness temperatures based on the thermal profile. The lower pressure levels probed by the Spitzer band compared to those by WFC3/G141 may contribute to the difference in period measurements between the two bands.}
  \label{fig:TP}
\end{figure}

Differential rotation and atmospheric circulation further complicate period measurements. GCM simulations in \citet{Showman2019} exhibit significant time-variable vertical shear in the zonal wind. As shown in Figure~\ref{fig:TP}, the pressure levels probed by the WFC3/G141 observations are deeper than those probed by the Spitzer \SI{4.5}{\micro\meter} channel. A vertical wind shear in \vhs  could lead to different period measurements between these two bands. Furthermore, GCMs in \citet{Showman2019} and \citet{Tan2021} predicted quasi-periodic velocity and directional changes in the zonal waves. Using periodograms of the GCM-simulated light curves, \citet{Tan2021} showed that the zonal waves could shift and broaden the power spectrum peaks near the underlying rotation periods. As a result, light curve periodicity indeed probes the rotation rate, but imprecisely. In the same periodogram, secondary peaks of zonal wavenumber $k=2$ are a common structure.

Assuming that the best-fitting sine waves probe atmospheric structures (e.g., planetary scale waves, jets, etc. See \S\ref{sec:map}), the shorter periods of Waves 1 and 2 relative to the Spitzer period suggest that they trace eastward-propagating structures and the period of Wave 3 is consistent with a $k=2$ harmonic. We convert periods to equatorial spin velocities by assuming $R=1.17\,R_{\mathrm{Jup}}$ \citep{Dupuy2020} and find that Waves 1 and 2 travel at \SI{1.1}{\kilo\meter\per\second} and \SI{3.0}{\kilo\meter\per\second}, respectively.  These results are at best order-of-magnitude estimates and are likely dominated by the observing-window-related systematics. Continuous and long-term time-resolved observations \citep[e.g.,][]{Apai2017,Apai2021} will allow  more accurate period measurements and wind speed constraints.

\subsection{Atmospheric Circulation in \vhs}
\label{sec:map}

\begin{figure*}[th]
  \centering
  \includegraphics[width=0.49\textwidth]{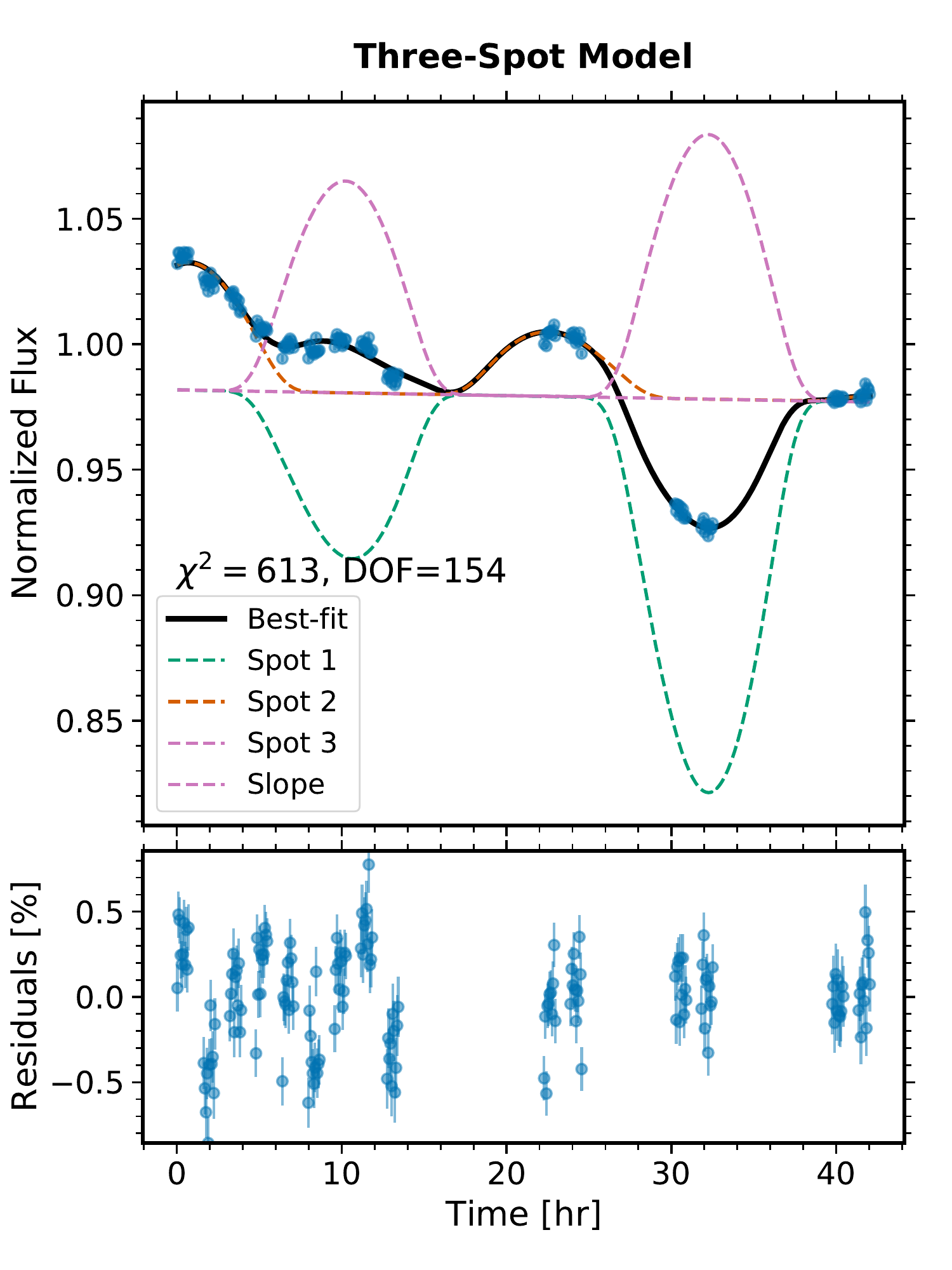}
  \includegraphics[width=0.49\textwidth]{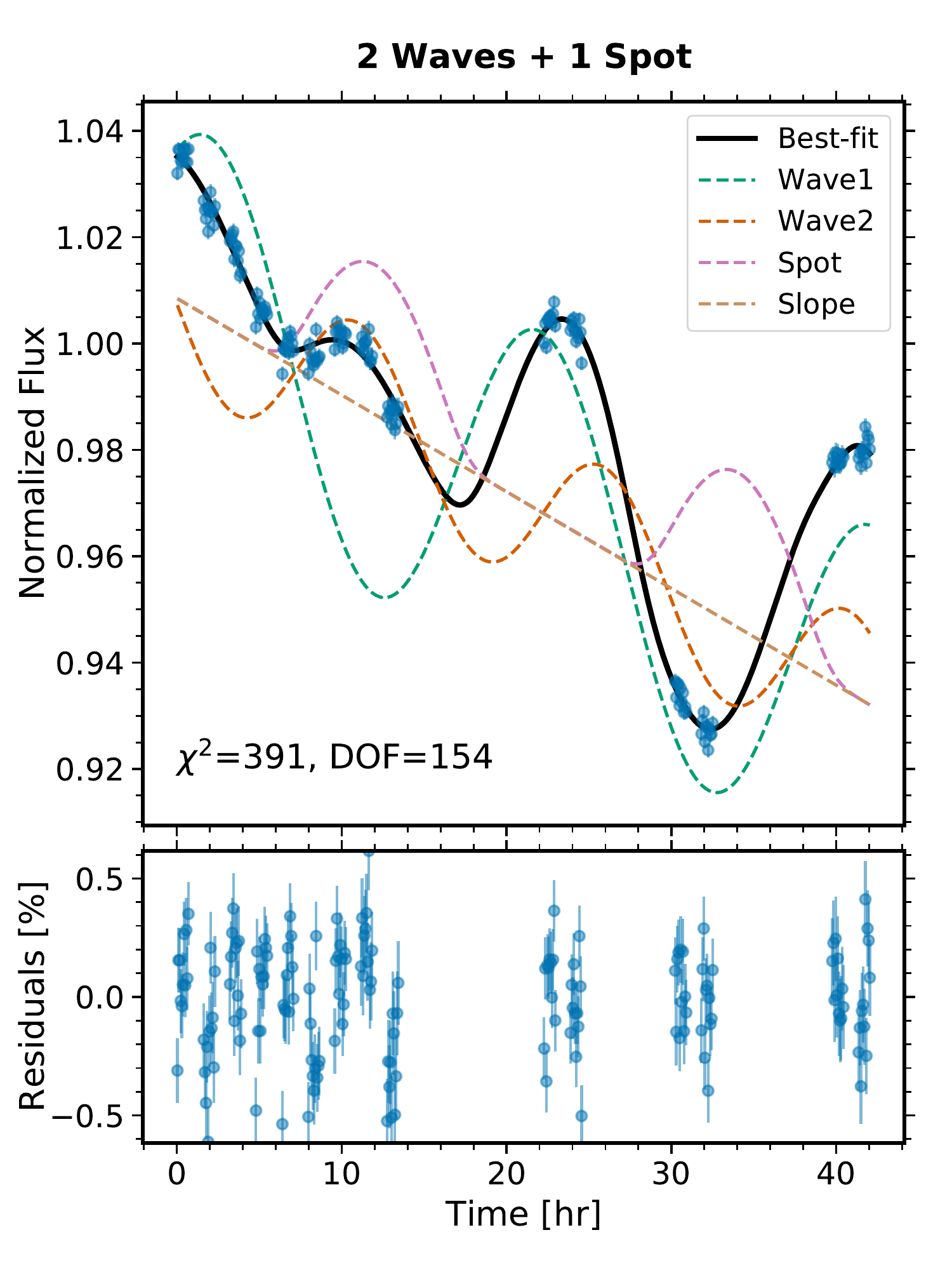}  
  \caption{Alternative de-compositions of the broadband light curves. Left: a model composed of three variable spots; right: a model composed of two waves and a variable bright spot. The solid lines, representing the best-fitting models, fit the data reasonably well. The dashed lines show individual model components. The fitting solutions are not unique due to parameter degeneracy. We show additional examples in Appendix \ref{sec:appendix}.}
  \label{fig:spotfit}
\end{figure*}

The markedly evolving light curves indicate vigorous atmospheric dynamical processes in \vhs. The underlying circulation patterns can be probed through top-of-the-atmosphere mapping that translates light curves into two-dimensional structures \citep[e.g.,][]{Crossfield2014,Karalidi2015,Plummer2022}. However, because our data are scarcely sampled and has a limited time baseline, the mapping results are highly degenerate.  In this subsection, we construct simplified mapping models that are plausible for \vhs based on recent theoretical work \citep[e.g.,][]{Zhang2014,Showman2019,Tan2021} and compare these models with the observed light curves.

Using GCMs that couple atmospheric dynamics with cloud formation and its radiative effect, \citet{Tan2021} identified two types of structures that cause light curve to evolve: zonal waves and vortices. Waves propagate zonally at various velocities and their changing phase differences result in emerging flux variability. Vortices are zonally stationary and impact the light curves by their stochastic transformations accompanied by cloud formation, dissipation, and thermal profile perturbation. Waves often have a stronger effect in the light curves, particularly in fast-rotating models ($P_{\mathrm{rot}}\le10$~hr) that are viewed equator-on \citep{Tan2021}. However, vortex variability is traceable in synthetic light curves of slow-rotating models ($P_{\mathrm{rot}}<10$~hr), because the vortex size increases linearly with the rotation period and long periods allow vortices to evolve sufficiently in one rotation. \vhs is a slow rotator, so both waves and vortices should be considered. Therefore, we adopt an agnostic approach and experiment with various combinations of waves and vortices. Our mapping models can be grouped into three scenarios: a) wave-dominated; b) spot-dominated; c) a mixture of wave and spots.

We briefly summarize the modeling approach here and provide details in Appendix~\ref{sec:appendix}. Zonal waves are modeled as sinusoids with periods being free parameters to allow for the possibility of high wavenumber components and differential rotation. Vortices are modeled as circular spots characterized by their sizes, positions, and contrasts. To reduce degeneracy and enable meaningful comparisons between data and the model, we place the following restrictions on spot properties:
\begin{enumerate}
\item The angular size of any spot is fixed to $15^{\circ}$, a limit determined by the Rossby deformation radius ($R_{\mathrm{spot}}\approx0.25R_{\mathrm{Jup}}$, Equation~\ref{eq:rossby}). Notably, because the spot size is a small fraction ($<10\%$) of a hemisphere, the light curve of a dark/bright spot transit appears like a eclipse/bump rather than a sinusoid.
\item The spots are longitudinally stationary and co-rotate with the sphere at a period of $P_{\mathrm{rot}}=22.0$~hr. Their latitudes are fixed to $45^{\circ}$ to reflect the fact that vortices are populated at mid-to-high latitude \citep{Tan2021}.
\item The spot has a uniform brightness contrast that can linearly vary with time, but the sign of contrast cannot change (i.e., a dark spot cannot become a bright one and \emph{vice versa}).
\end{enumerate}
With these constraints, a spot is modeled by three parameters: longitude, average contrast, and contrast variability. The three model types are constructed by linearly combining sinusoids and spot light curves.

\subsubsection{Scenario 1: Wave-Dominated Mapping}

Assuming the light curve evolution is dominated by zonal waves, we can map the sinusoids in the three-sine-wave model (Section~\ref{sec:lc-fit}) into three atmospheric waves. This is similar to the planetary-scale wave mapping that explained long-term ($>100 P_{\mathrm{rot}}$ ) photometric evolution in several brown dwarfs \citep{Apai2017,Apai2021}. Assuming the true rotation period is the 22.0~hr period measured by Spitzer \citep{Zhou2020a}, the 18.6~hr and 15.1~hr sinusoids (Waves 1 and 2) are two eastward traveling waves in which cloud thickness modulates in a wavenumber $k=1$ pattern and the 10.5~hr sinusoid corresponds (Wave 3) to thermal profile modulation in a $k=2$ pattern. Differential propagation of the three waves can explain most of the photometric and spectroscopic variability observed in Epoch 2020 except the linear trend. A polar spot in which cloud dissipates in combination with a slightly tilted spin-axis that allows the polar region to be constantly visible is a possible explanation.

\subsubsection{Scenario 2: Spot-dominated Mapping}

We can also begin with the assumption that the observed light curve evolution is the consequence of vortex transformation and construct a light curve model that only contains time-varying circular spots and a long-term linear trend. We fit this model to the Epoch 2020 light curve with an incremental increase in the number of spots. We find that a reasonable fit requires at least three spots ($\chi^{2}=613$, DOF$=154$, in comparison to $\chi^{2}=346$, DOF$=154$ for the three-sine-wave model). The left panel of Figure~\ref{fig:spotfit} shows a possible solution consisting of a dimming bright spot and a pair of bright and dark spots locating at the opposite longitudes. Fitting to the light curve requires the latter spot pairs to change contrasts in divergent directions. All three spots have significant contrast variations between the two periods. The contrasts of the two bright spots change by 25\% and 50\%, corresponding to brightness temperature variations ($\Delta T_{\mathrm{B}}$) of 60~K and 110~K, respectively. Variability introduced by the dark spot (8\% to 18\%) exceeds the maximum for a $R=15^{\circ}$ spot (${\sim}6.6\%$ for a non-emitting spot), suggesting that a size much greater than the Rossby deformation radius or multiple co-evolving dark spots are required for this solution to be physically plausible. When the sign of spot contrast is allowed to change, a two-spot model can fit the light curve, albeit an even more rapid contrast variation is needed (Figure~\ref{fig:2spot}).

\subsubsection{Scenario 3: A mixture of Waves and Spots}

Finally, we attempt to reproduce the light curve with a mixture of waves and spots. The number of mapping elements is limited to three or fewer to avoid over-fitting. The right panel of Figure~\ref{fig:spotfit} shows the best-fitting configuration ($\chi^{2}=391$, DOF$=154$, in comparison to $\chi^{2}=346$, DOF$=154$ for the three-sine-wave model). This model consists of two sinusoids ($P_{1}=20.0\pm0.1$~hr, $A_{1}=3.4\pm0.1\%$; $P_{2}=14.9\pm0.1$~hr, $A_{2}=1.5\pm0.1\%$) and a bright spot with a nearly constant contrast (60\% brighter or $\Delta T_{\mathrm{B}}=125$~K higher than the background). The solution is not unique and alternative model configurations (e.g., one sinusoid and two spots) can fit the light curve reasonably well (see examples in Appendix~\ref{sec:appendix}).

\subsubsection{Model Interpretation and Future Perspectives}


Models that fit the light curve well all contain features that are not fully consistent with atmospheric dynamical models. For example, the wave-dominated mapping requires differential rotation of high velocity  ($v\sim2$ km\,s$^{-1}$) that exceeds the maximum wind speed in the GCMs of \citet{Tan2021}. In the vortex-dominated case, the rapid and seemingly coordinated variations of two spots in the opposite longitudinal positions are also improbable. Increasing the model flexibility by adding more spots or allowing the spot size to vary with time may reconcile these incompatibilities, although these more complex models are not warranted by our data. Meanwhile, the extremely high-amplitude variability seen in \vhs poses a challenge to state-of-the-art GCMs that do not produce variability of more than a few percent. More realistic cloud and radiative transfer modeling could result in a better match between GCMs and the observed results.

Although we purposely restrict our models to several simplified cases, we still find a variety of maps that can reproduce \vhs's light curve (Appendix~\ref{sec:appendix}). The high degeneracy underscores the limitation of short and sparsely sampled observations in probing circulation patterns. Long-term and contiguous light curves can distinguish the origin of light curve evolution, because the difference between wave beating and stochastic vortex variability becomes unambiguous when the time baseline is beyond one or two rotation periods. For all published long-term (${>}30$ rotations) brown dwarf light curves (2M1324, $P_{\mathrm{rot}}=13.2$~hr; 2M2139, $P_{\mathrm{rot}}=8.2$~hr; and SIMP0136, $P_{\mathrm{rot}}=2.4$~hr in \citealt{Apai2017} and Luhman16 AB, $P_{\mathrm{rot}}=5.3$~hr\footnote{This is the period of Luhman16 B that shows a higher-amplitude in the binary.} in \citealt{Apai2021}), zonal waves dominated maps with occasional additions of spots are the only model that have been shown to agree with the observations. Whether \vhs, a slower rotator, has similar atmospheric dynamical properties requires further investigation. Doppler imaging \citep{Crossfield2014} and time-resolved polarization observations \citep[e.g.,][]{Millar-Blanchaer2020,Mukherjee2021} can also offer unique constraints on the circulation patterns of \vhs.

  \subsection{The Extremely High-Amplitude Variability of \vhs}
  The brightness change observed in \vhs is the highest found in brown dwarfs. The 33\% peak-to-peak flux variation observed in \vhs exceeds the previous largest brown dwarf variability amplitude observed in the $J$-band light curve of 2MASS J21392676+0220226 \citep{Radigan2012,Apai2013} In \citet{Zhou2020a}, we have shown that the asynchronously observed spectral variability in the HST/WFC3/G141 band and the Spitzer \SI{4.5}{\micro\meter} channel could be very well explained by cloud optical depth changes. In this work, we further demonstrate that thermal profile anomalies in addition to heterogeneous clouds may contribute to observed spectral variability.

  \vhs has several properties that are favorable for producing high-amplitude variability. It is likely young and has a low surface gravity \citep{Dupuy2020}, resulting in a large atmospheric scale height and geometrically thick clouds. The latter is also evident from its unusually red infrared colors \citep{Gauza2015,Rich2016}. These conditions permit significant spatial variations in cloud optical depth that leads to high-amplitude rotational modulations. \vhs has an edge-on viewing geometry that maximizes the effect of heterogeneous atmospheric structures on photometric and spectral variability \citep{Vos2017}. Results from \citet{Tan2020b,Tan2021} also show that the relatively slow rotation rate of \vhs may also indirectly contribute to its high variability, because slow rotators are likely to have thicker clouds, larger-sized vortices and therefore larger temporal flux changes.

  Brown dwarfs and planetary-mass objects that share almost identical near-infrared spectra to \vhs,  such as PSO~J318.5-22 \citep{Biller2015} and WISEP~J004701+680352 \citep{Lew2016}, also exhibit high-amplitude variability. Directly imaged exoplanets with similarly red near-infrared colors (e.g., HR~8799~bcde \citealt{Marois2008a,Marois2010}, HD~95086~b \citealt{Rameau2013,derosa2016}) are candidates for detecting variable brightness and spectra. Based on the complexity of heterogeneous structures in \vhs we have found in this work, monitoring programs for these planets \citep[e.g.,][]{Apai2016,Biller2021,Wang2022} will provide a better understanding of how these physical properties play a role in the observed rotational variability.

  \section{Summary}
  \label{sec:conclusions}

  We analyzed HST/WFC3 time-resolved spectra of the L7-type planetary-mass companion \vhs and found high-amplitude and wavelength-dependent variability. The evolving spectra reveal heterogeneous distributions of clouds and thermal profiles in the atmosphere of the companion and enable an investigation of the atmospheric dynamical processes that shape these structures. We list our findings as follows:
  \begin{enumerate}
\item In a new HST time-resolved observing campaign, \vhs was spectroscopically monitored by WFC3/G141 with a time baseline of 42 hr (approximately two rotation periods). The brown dwarf companion exhibited significant variability in its spectrum and light curve between the two rotation periods.

\item \vhs's fast-evolving light curves are more complex than previously reported. A combination of three sine waves in which the periods are free parameters fits the light curve well. For the \SIrange{1.12}{1.65}{\micro\meter} broadband light curve, the best-fitting sine-wave periods are $18.8\pm0.2$~\si{\hour}, $15.1\pm0.2$~\si{\hour}, and $10.6\pm0.1$~\si{hour}, and the corresponding peak-to-peak amplitudes are $5.8\pm0.8$\si{\percent}, $4.6\pm0.7$\si{\percent}, and $1.4\pm0.1$\si{\percent}. A linear trend is a necessary component of the light curve model to explain the long-term light curve variation.

  \item \vhs exhibits clear spectral variability in the new epoch. The difference between the maximum and minimum brightness spectra decreases in the \SI{1.4}{\micro\meter} band, which is qualitatively consistent with its behavior in a previous HST observing epoch \citep{Bowler2020,Zhou2020a}.
  
\item  We fit the three-sine-wave model to spectroscopically resolved light curves to quantify the spectral variability. We find that the amplitudes of all three waves and the linear trend are wavelength-dependent.  The two long-period waves and the linear trend have a higher amplitude in the continuum than in the \water band, resembling the spectral variability caused by clouds with heterogeneous optical depths. In contrast, the shortest period wave has a higher amplitude in the water band than in the continuum, resembling spectral variability caused by perturbations to the thermal profile.

\item By consistently reducing two epochs of HST/WFC3 observations of \vhs, we estimate \vhs's variability on a time-scale of nearly 900 rotation periods and find that the variability amplitude between the two epoch is greater than those within individual epochs. The peak-to-peak flux difference is $33.3\pm2.2\%$ in the G141 broadband and reaches to $37.6\pm2.2\%$ at \SI{1.27}{\micro\meter}. These brightness changes are among the strongest ever found in brown dwarfs.

\item The spectral change between the two epochs has a weak wavelength dependence. The decrease of variability amplitude in the \SI{1.4}{\micro\meter} band detected in individual epochs is absent in the cross-epoch measurement. This suggests different atmospheric structures are responsible for the long-term variability  than those for variability on the rotational timescale.

  \item The rapidly evolving light curves can be reproduced by a variety of physically plausible models that are constructed based on zonal waves, spots, or a mixture of waves and spots. Long-term and continuous light curves can help pinpoint the circulation regime of \vhs.

\item Uncertainties in the true underlying light curve model can impair the degree to which we can estimate \vhs's rotation period. Atmospheric evolution can  bias the period results by 5 to 10\%, far exceeding the least-squares-determined uncertainties. Reducing these systematic uncertainties also requires extending the observing time baseline.
  \end{enumerate}

 Our work has two broad implications that should be considered in future studies. First, the strong variability of \vhs, an excellent analog to giant exoplanets such as HR8799~bcde, further proves that atmospheric dynamics fundamentally shapes the spectral appearance of planetary atmosphere, and that understanding the 3D atmospheric dynamics is critical for interpreting time-series observations. Second, long-term spectroscopic monitoring observations deliver a wealth of information. Expanding such observing programs to more targets and broader wavelength ranges (e.g., with JWST) will provide valuable data that help probe atmospheric dynamics in planets and brown dwarfs.

 \vspace{2em}
 We thank anonymous referees for detailed and constructive reports. We thank Dr. Xianyu Tan for a helpful discussion about atmospheric dynamical modelings. Y.Z. acknowledges support from the Harlan J. Smith McDonald Observatory Fellowship and the Heising-Simons Foundation 51 Pegasi b Fellowship. B.P.B. acknowledges support from the National Science Foundation grant AST-1909209, NASA Exoplanet Research Program grant 20-XRP20$\_$2-0119, and the Alfred P. Sloan Foundation. C.V.M acknowledges support from the National Science Foundation grant AST-1910969. This research has made use of the NASA Exoplanet Archive, which is operated by the California Institute of Technology, under contract with the National Aeronautics and Space Administration under the Exoplanet Exploration Program.  The observations and data analysis works were supported by program HST-GO-16036. Supports  for  Program  numbers  HST-GO-16036 were provided by NASA through a grantfrom the Space Telescope Science Institute, which is operated by the Association of Universities for Research in Astronomy, Incorporated, under NASA contract NAS5-26555.

 \software{\texttt{Numpy} \citep{2020NumPy-Array}, \texttt{Scipy} \citep{2020SciPy-NMeth}, \texttt{Matplotlib} \citep{Hunter2007}, \texttt{Seaborn} \citep{Waskom2021}, \texttt{Astropy} \citep{Robitaille2013}, \texttt{Pysynphot} \citep{2013ascl.soft03023S}, \texttt{Lmfit} \citep{Lmfit}, \texttt{emcee} \citep{Foreman-Mackey2012},
 \texttt{Starry} \citep{Luger2019}}

\facility{Hubble Space Telescope}

\appendix
\restartappendixnumbering 
\section{Decomposing the Epoch 2020 Light Curve}
\label{sec:appendix}
\subsection{The modeling approach}
To explore the circulation patterns in VHS~1256~b, we decompose its Epoch 2020 broadband light curve into sinusoids (zonal waves) and circular spots (vortices). The light curve model is akin to the multi-sinusoidal model (Equation~\ref{eq:3-2}) with inclusions of spot components. A $N$ sinusoids and $M$ spots model is expressed as:
\begin{equation}
  \label{eq:a1}
  F (t) = C_{0} + C_{1}t + \sum_{i=1}^{N}\mathrm{sin}(P_{i}, A_{i}, B_{i}) + \sum_{j=1}^{M}\mathrm{spot}(l,\theta,V).
\end{equation}
$l$, $\theta$, and $V$ are the longitude, brightness contrast, and linear contrast variability of a spot. One sinusoid or one spot adds three free parameters, so the total number of free parameter is $2+3*(N+M)$. We restrict model complexity by $N+M\le3$.

The spot light curve is pre-calculated using \texttt{starry} \citep{Luger2019} with $R_{\mathrm{spot}}=15^{\circ}$, $\mathrm{lat_{\mathrm{spot}}}=45^{\circ}$, and a contrast ratio of 2 (i.e., the spot is 100\% brighter than the global average, corresponding to $\Delta T_{\mathrm{B}}=189$~K in a $T_{\mathrm{B}}=1000$~K sphere). The spot size is assumed to be the Rossby deformation radius, which is the typical size of a vortex driven by cloud radiative feed back \citep{Tan2021}
\begin{equation}
  \label{eq:rossby}
  R_{\mathrm{Rossby}}=\frac{c_{g}}{2\Omega \sin\phi} = 0.25 R_{\mathrm{Jup}}* \Bigl(\frac{c_{g}}{2000\,\mathrm{m\,s^{-1}}}\Bigr)
  \Bigl(\frac{T}{22\,\mathrm{hr}}\Bigr) \Bigl(\frac{\sin\phi}{\sin(\pi/4)}\Bigr)^{-1},
\end{equation}
in which $c_{g}$ is the gravity wave phase speed, $\Omega$ is the rotation rate, and $\phi$ is the spot latitude.
Given a set of $l$, $\theta$, and $V$, the spot light curve is phase shifted by $l$ and multiplied by a linear amplitude term $\theta + Vt$. In all but one cases (Figure~\ref{fig:2spot}), $\theta$ and $V$ are restricted \emph{a priori} to preclude sign change in $\theta + Vt$ during our $42$~hr observing window.

We optimize the free parameters by first a least-square fitting using \texttt{LMFIT}  and then a MCMC. The likelihood function and posterior probability are derived in the same way as the multi-sinusoidal model fit. In each case, the $\chi^{2}$ value is recorded to evaluate the fitting quality. Among all experiments with $N+M\le3$, the three-sine-wave model has the lowest $\chi^{2}$. Nevertheless, relaxing the restrictions and then fine-tuning the spot properties may result in even better fits.

For each set of $N$ and $M$, multiple solutions that corresponds to a local $\chi^{2}$ minimum may exist. We do not attempt to exhaust all possible solutions, but only use these experiments to demonstrate the fact that our current data is not yet able to pinpoint the circulation patterns in \vhs. 

\subsection{Examples of Light Curve Decomposition}
\label{sec:a12}

\begin{figure*}[th]
  \centering
  \includegraphics[width=0.4\textwidth]{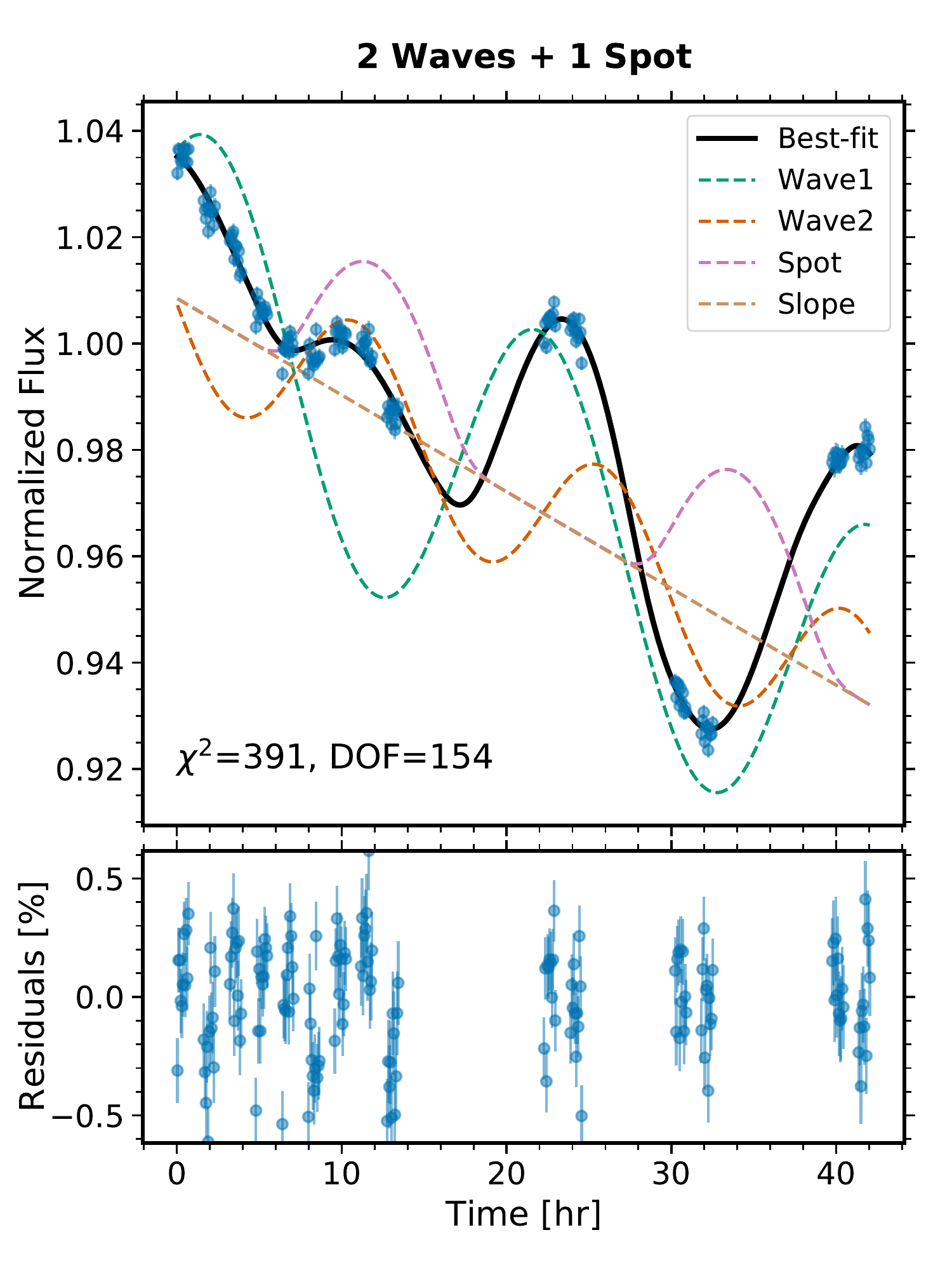}
  \includegraphics[width=0.4\textwidth]{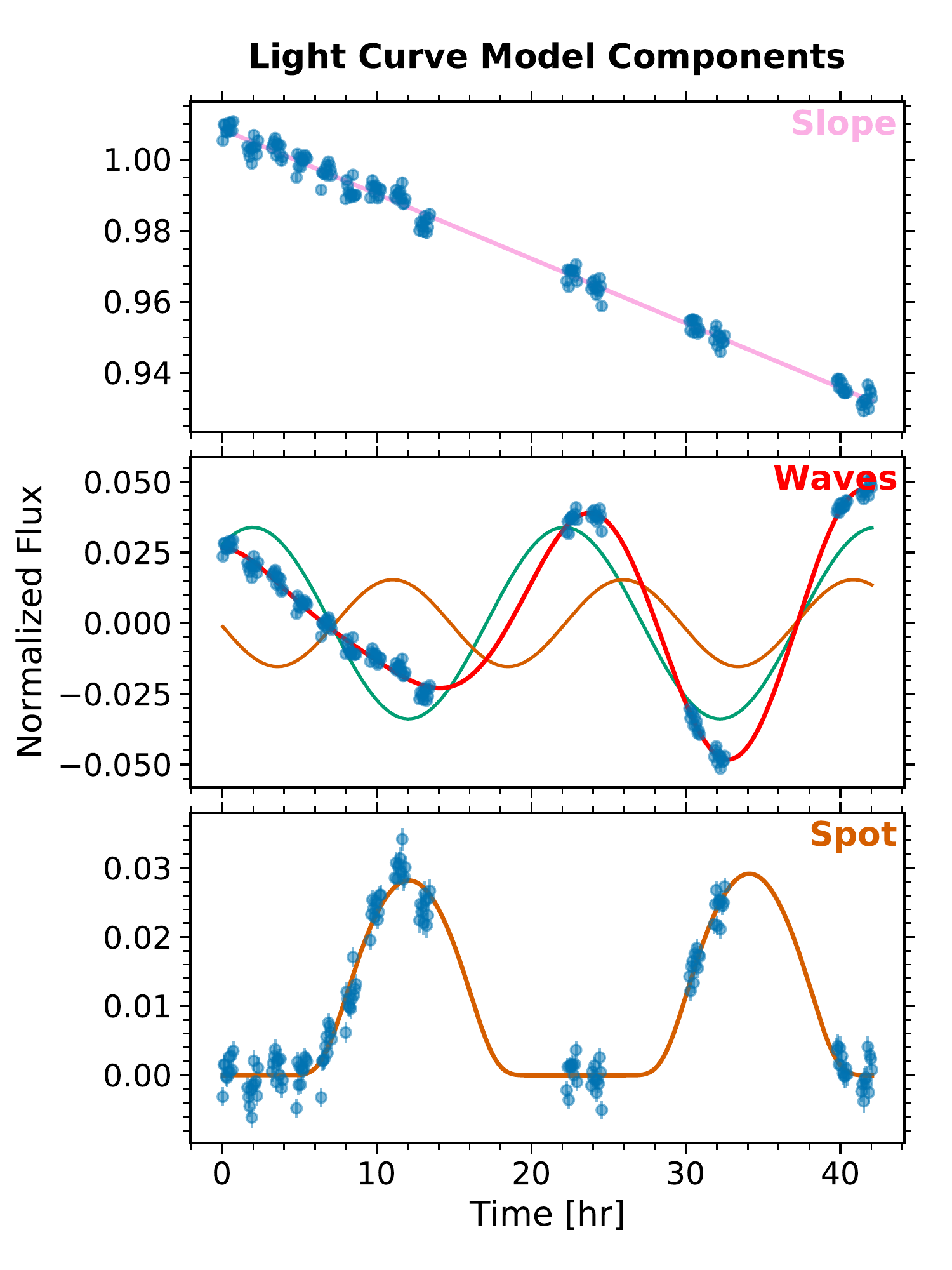}
  \includegraphics[height=0.5\textheight]{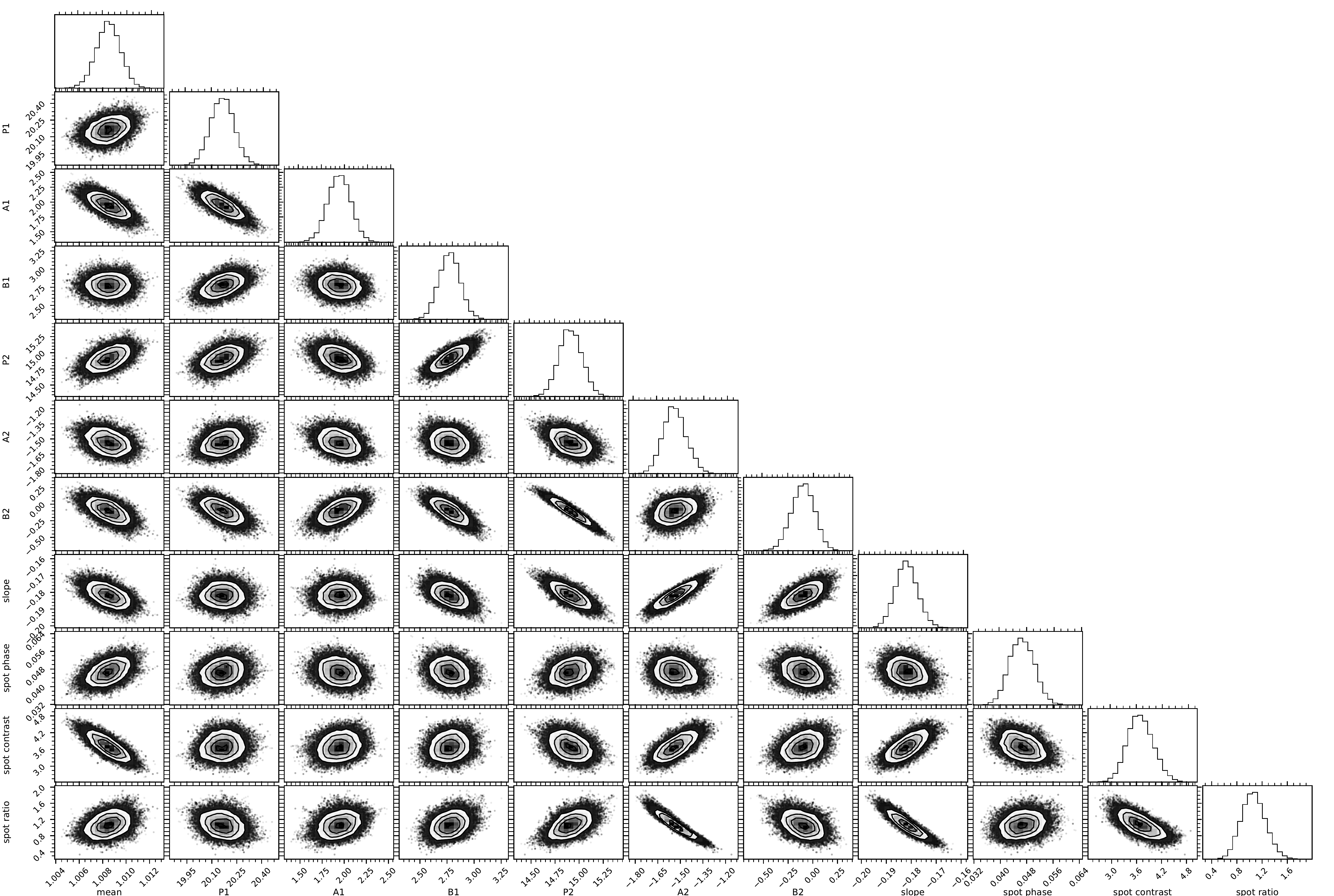}
  \caption{Two waves + one spot light curve decomposition.}
  \label{fig:2wave_1spot}
\end{figure*}

\begin{figure*}[th]
  \centering
  \includegraphics[width=0.4\textwidth]{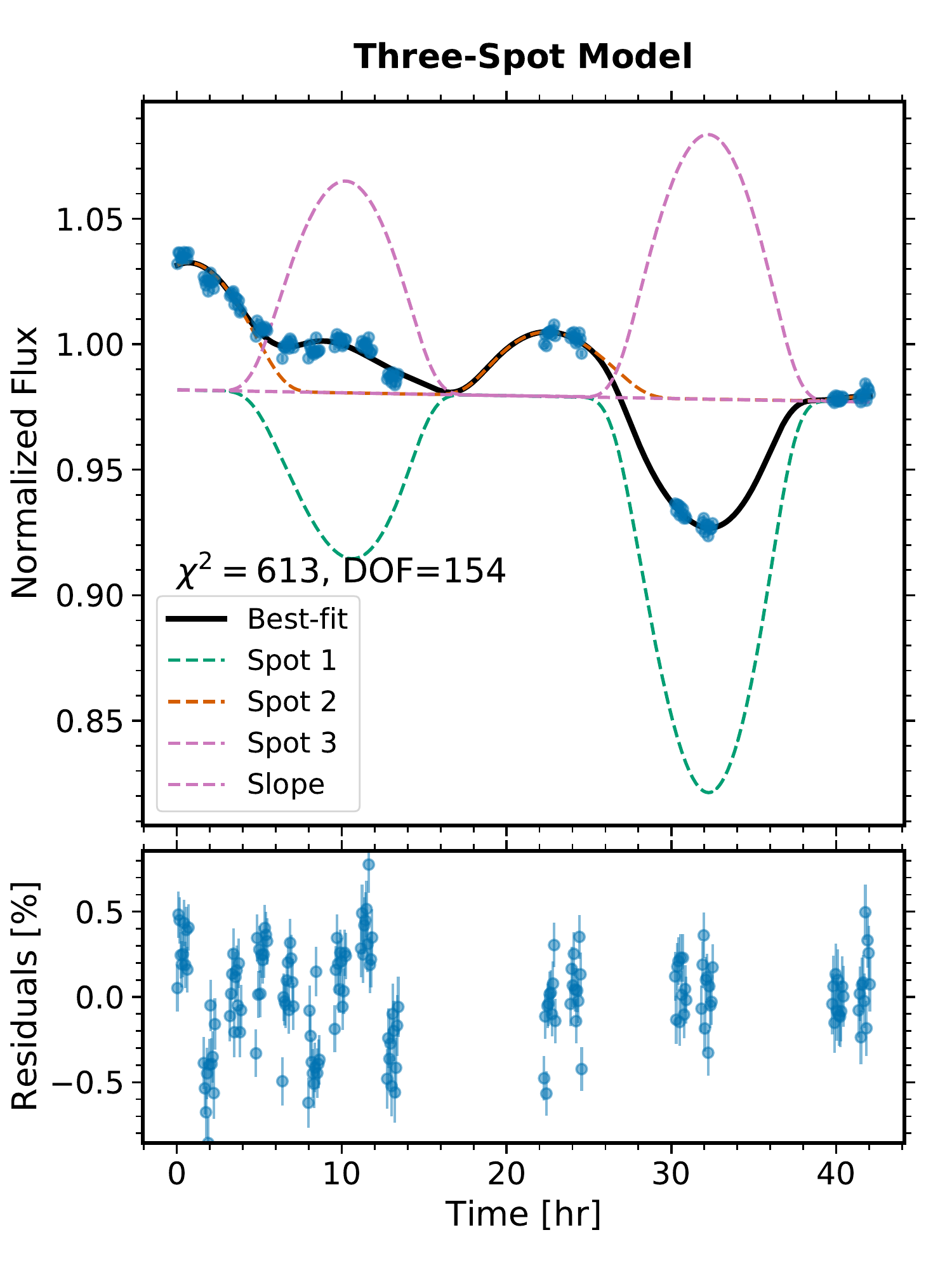}
  \includegraphics[width=0.4\textwidth]{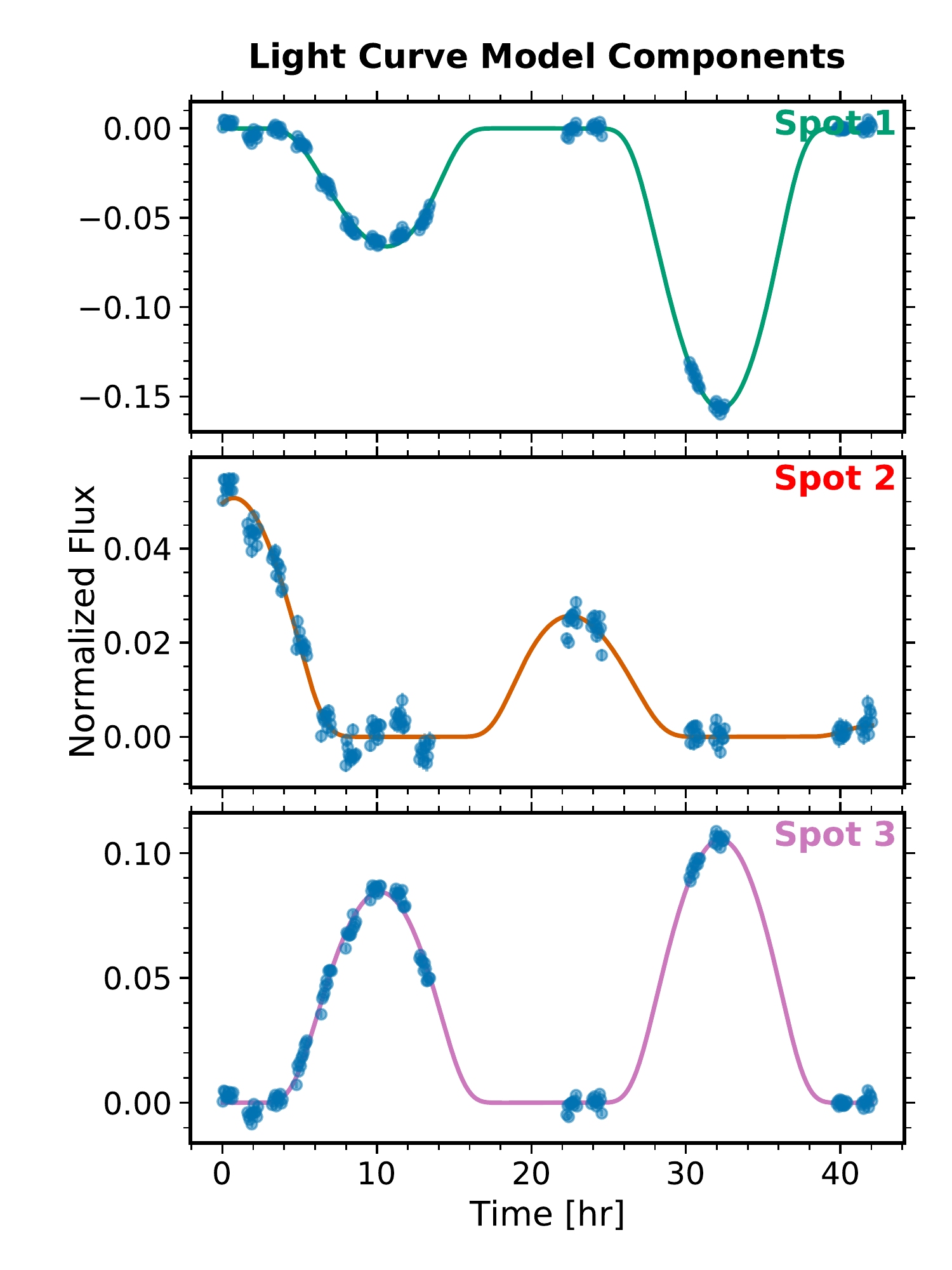}
  \includegraphics[height=0.5\textheight]{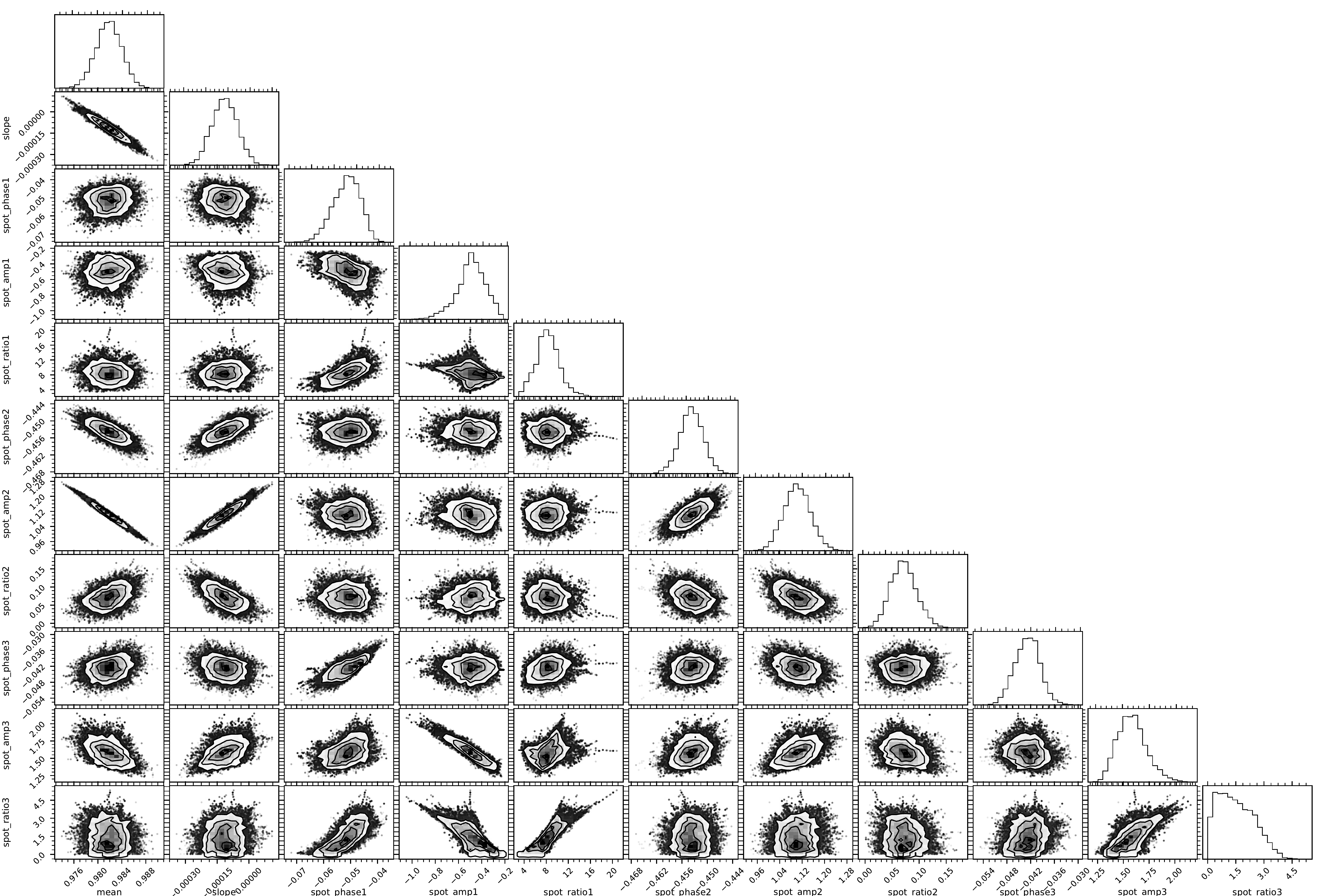}
  \caption{Three-spot light curve decomposition.}
  \label{fig:3spot}
\end{figure*}

\begin{figure*}[th]
  \centering
  \includegraphics[width=0.4\textwidth]{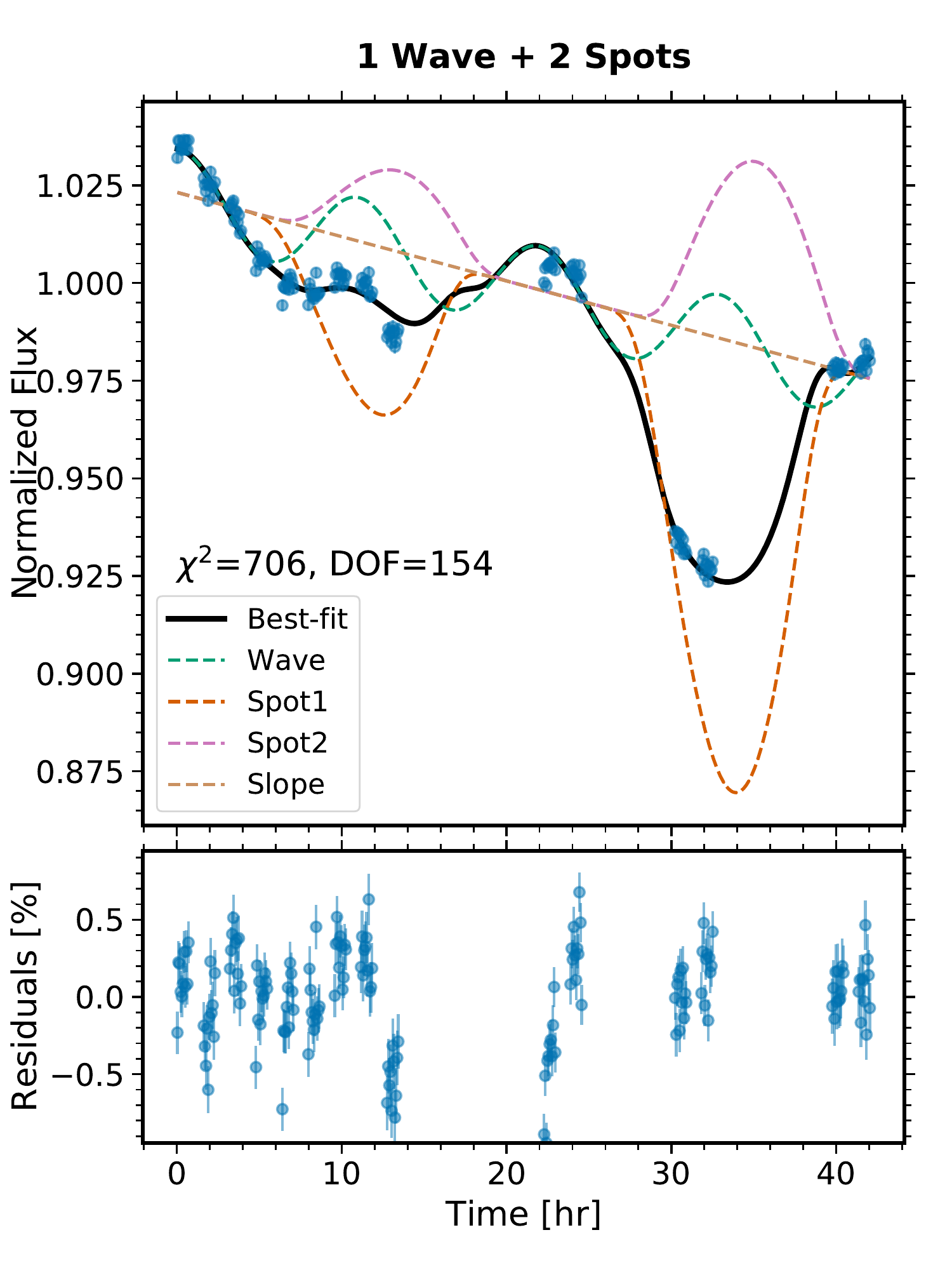}
  \includegraphics[width=0.4\textwidth]{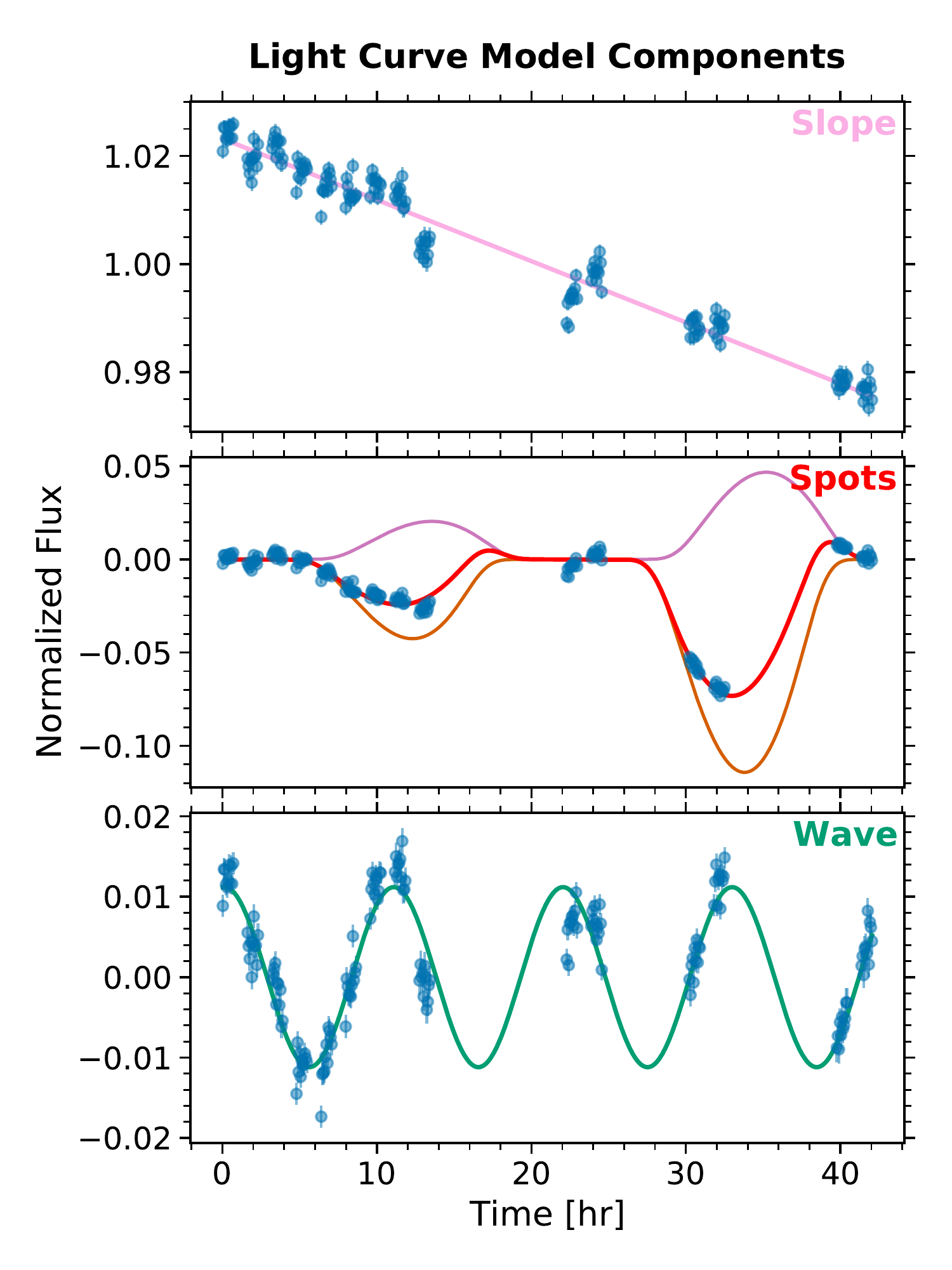}
  \includegraphics[height=0.5\textheight]{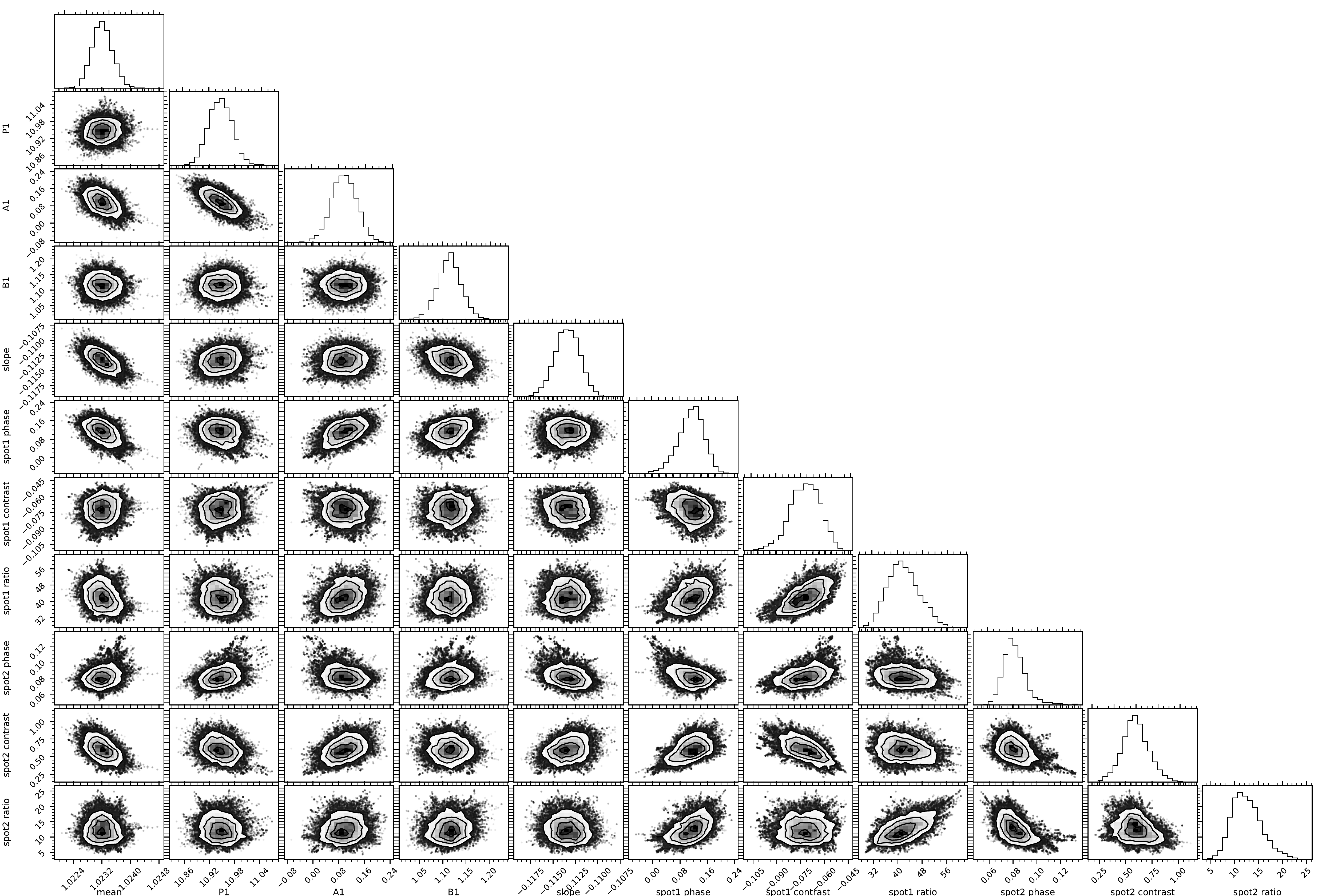}
  \caption{One wave ($k=2$) + two spots light curve decomposition.}
  \label{fig:1wave_2spot}
\end{figure*}

\begin{figure*}[th]
  \centering
  \includegraphics[width=0.4\textwidth]{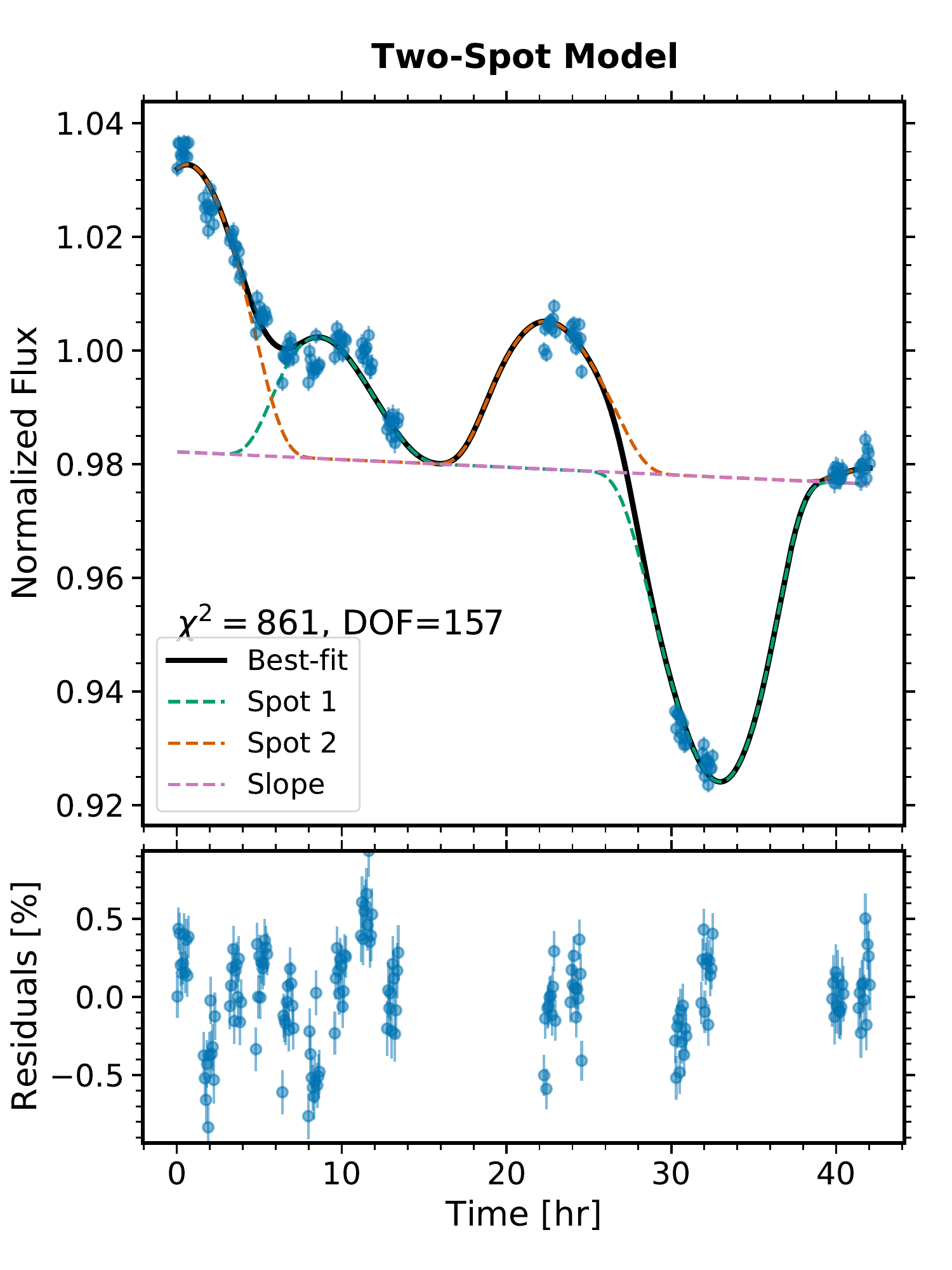}
  \includegraphics[width=0.4\textwidth]{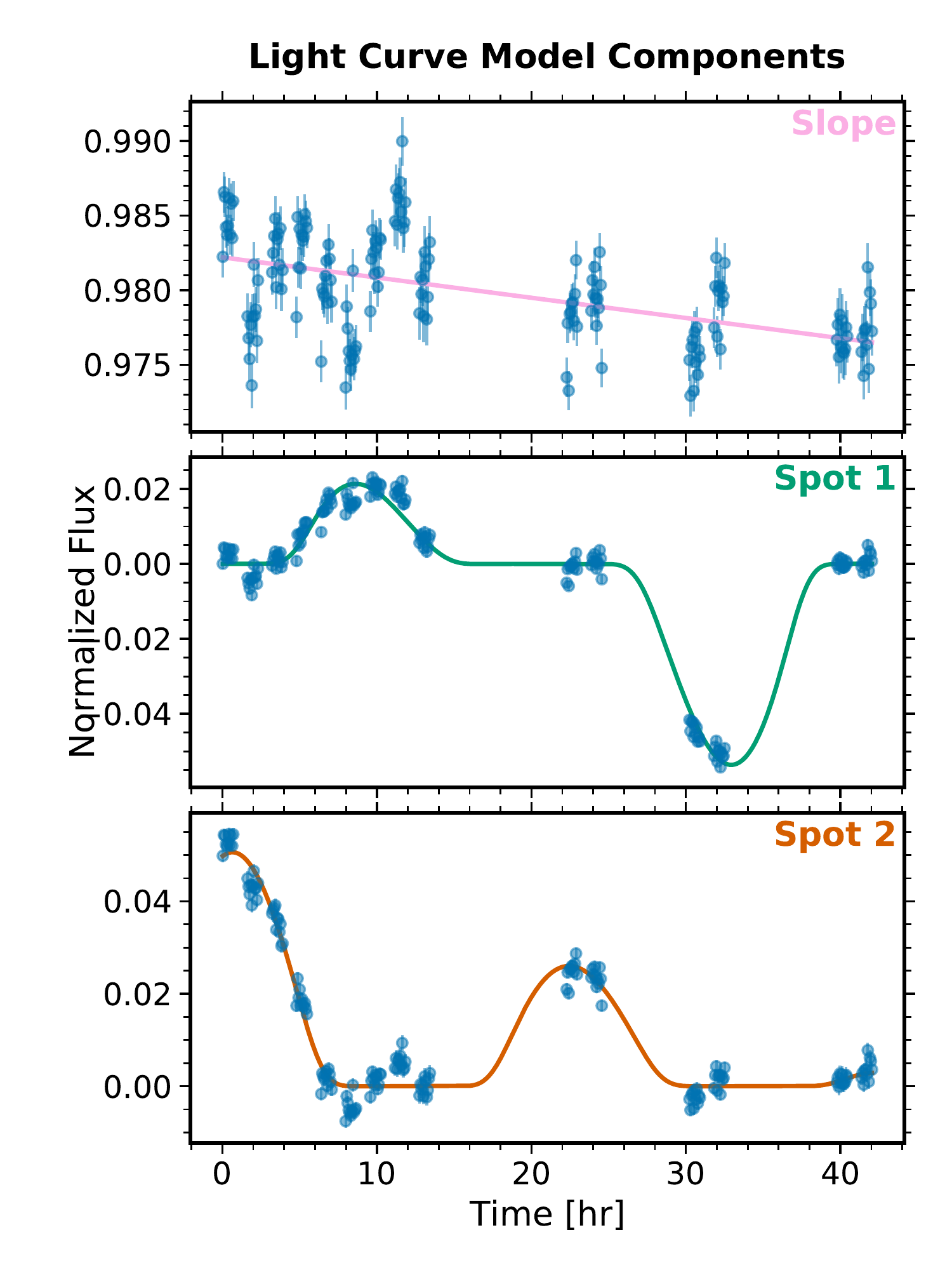}
  \includegraphics[height=0.5\textheight]{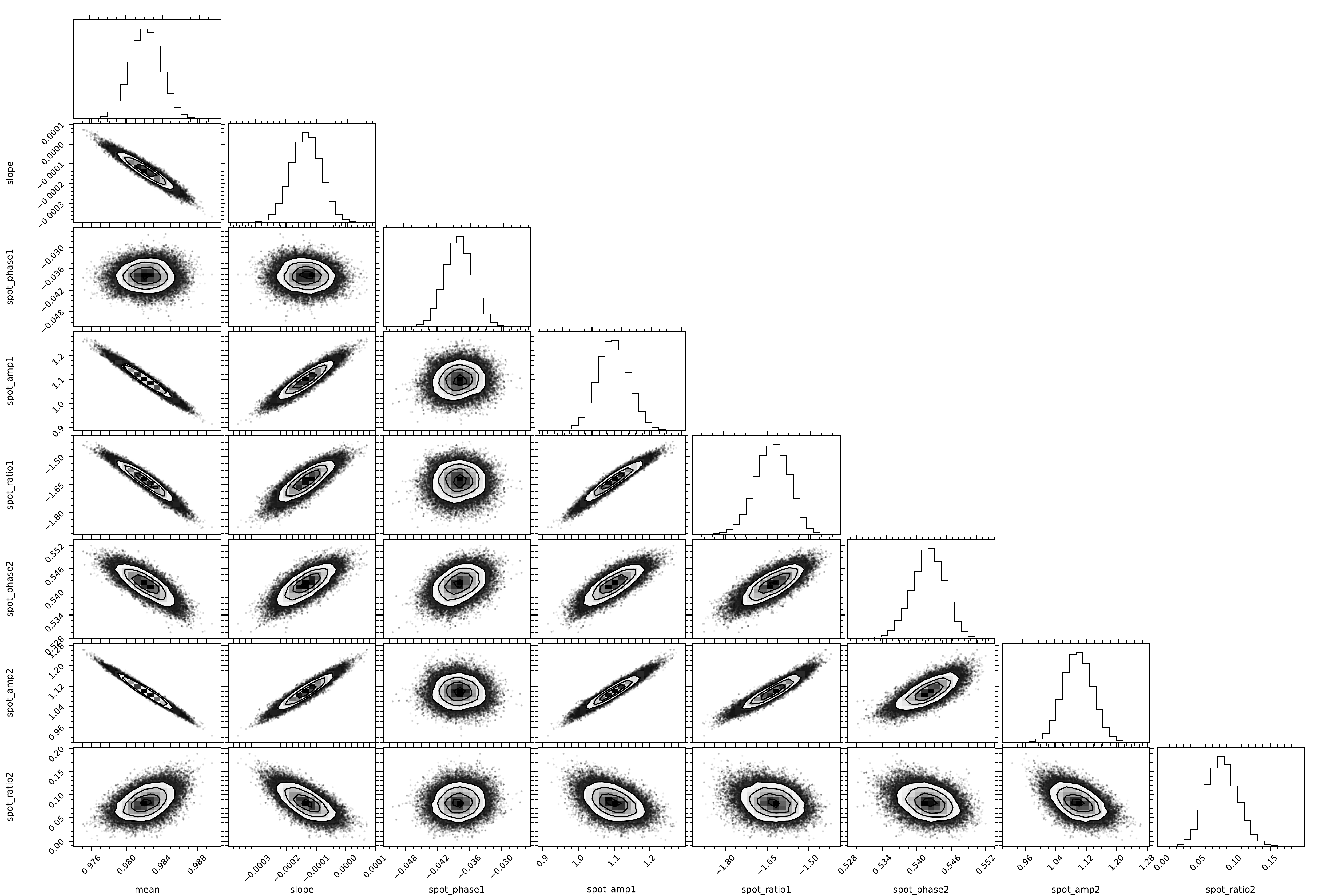}
  \caption{Two-spot light curve decomposition.}
  \label{fig:2spot}
\end{figure*}

Figures \ref{fig:2wave_1spot} to \ref{fig:2spot} illustrate a few examples of light curve decomposition results. These four figures correspond to: \\
\hspace*{0.2in}\ref{fig:2wave_1spot}, two sinusoids and one bright spot;\\
\hspace*{0.2in}\ref{fig:3spot}, one dark spot and two bright spots;\\ \hspace*{0.2in}\ref{fig:1wave_2spot}, one ($k=2$) wave, one bright spot, and one dark spot;\\
\hspace*{0.2in}\ref{fig:2spot}, two spots. In Figure~\ref{fig:2spot}, the spot contrast is allowed to switch sign. The rapid spot evolution enables a good fit with only two spots.

In each Figure, the upper panel is in the same format as Figure~\ref{fig:lcfit}, showing the observation-model comparison on the left and highlighting model components on the right. The lower panel is a corner plot demonstrating the posterior distributions of model parameters.

\end{document}